\documentstyle[12pt,bezier]{article}
\def\emline#1#2#3#4#5#6{%
       \put(#1,#2){\special{em:moveto}}
       \put(#4,#5){\special{em:lineto}}}
\def\newpic#1{}
\catcode`\@=11
\def\draftlabel#1{{\@bsphack\if@filesw {\let\thepage\relax
   \xdef\@gtempa{\write\@auxout{\string
      \newlabel{#1}{{\@currentlabel}{\thepage}}}}}\@gtempa
   \if@nobreak \ifvmode\nobreak\fi\fi\fi\@esphack}
        \gdef\@eqnlabel{#1}}
\def\@eqnlabel{}
\def\@vacuum{}
\def\draftmarginnote#1{\marginpar{\raggedright\scriptsize\tt#1}}
\def\draft{\oddsidemargin -.5truein
        \def\@oddfoot{\sl preliminary draft \hfil
        \rm\thepage\hfil\sl\today\quad\militarytime}
        \let\@evenfoot\@oddfoot \overfullrule 3pt
        \let\label=\draftlabel
        \let\marginnote=\draftmarginnote
@
 \def\@eqnnum{(\theequation)\rlap{\kern\marginparsep\tt\@eqnlabel}%
\global\let\@eqnlabel\@vacuum}  }

\def\numberbysection{\@addtoreset{equation}{section}
        \def\theequation{\thesection.\arabic{equation}}}

\def\underline#1{\relax\ifmmode\@@underline#1\else
        $\@@underline{\hbox{#1}}$\relax\fi}

\catcode`@=12
\relax
\numberbysection

\newcommand{\Aol}{A_0^{(l)}}
\newcommand{\Ail}{A_1^{(l)}}
\newcommand{\Bl}{{B}^{(l)}}
\newcommand{\Blis}{\left[ \left( B^{(l)} \right)^{-1} \right]}
\newcommand{\Btl}{\tilde{B}^{(l)}}
\newcommand{\Dl}{{D}^{(l)}}
\newcommand{\Dtl}{\tilde{D}^{(l)}}
\newcommand{\re}{{\rm e}}

\newcommand{\rJ}{{\rm J}}
\newcommand{\rJt}{{\rm J}^{\dagger}}
\newcommand{\rJo}{{\rm J}_{0}}
\newcommand{\rJot}{{\rm J}_{0}^{\dagger}}
\newcommand{\St}{S^{\dagger}}

\newcommand{\bB}{{\bf B}}
\newcommand{\bBt}{{\bf B}^{\dagger}}

\newcommand{\bI}{{\bf I}}
\newcommand{\bJ}{{\bf J}}
\newcommand{\bJo}{{\bf J}_0}
\newcommand{\bJot}{{\bf J}_0^\dagger}
\newcommand{\bJt}{{\bf J}^{\dagger}}

\newcommand{\bt}{{\bf t}}
\newcommand{\bR}{{\bf R}}
\newcommand{\bRo}{{\bf R}_{0}}

\newcommand{\bs}{{\bf s}}
\newcommand{\bv}{{\bf v}}
\newcommand{\cA}{{\cal A}}

\newcommand{\cE}{{\cal E}}

\newcommand{\cG}{{\cal G}}
\newcommand{\cH}{{\cal H}}

\newcommand{\cP}{{\cal P}}
\newcommand{\cQ}{{\cal Q}}

\newcommand{\cT}{{\cal T}}

\newcommand{\hbI}{\hat{\bf I}}
\newcommand{\hbIo}{\hat{\bf I}_0}
\newcommand{\hIo}{\hat{I}_0}
\newcommand{\hIi}{\hat{I}_1}
\newcommand{\hk}{\hat{k}}
\newcommand{\hp}{\hat{p}}
\newcommand{\hP}{\hat{P}}

\newcommand{\Sls}{\left[ S_l \right]}
\newcommand{\Slis}{\left[ S_l^{-1} \right]}

\newcommand{\tL}{\tilde{L}}
\newcommand{\un}{^{(n)}}

\newcommand{\Xl}{{\bf X}^{(l)}}
\newcommand{\Xol}{{\bf X}_0^{(l)}}
\newcommand{\Xil}{{\bf X}_1^{(l)}}
\newcommand{\an}{\alpha}
\newcommand{\bn}{\beta}

\newcommand{\ad}{_{\alpha}}

\newcommand{\ajd}{_{\alpha,j}}
\newcommand{\adl}{_{\alpha,l}}
\newcommand{\bjd}{_{\beta,j}}

\newcommand{\bd}{_{\beta}}

\newcommand{\abd}{_{\alpha\beta}}

\newcommand{\au}{^{(\alpha)}}

\newcommand{\aut}{^{(\alpha)\dagger}}

\newcommand{\aju}{^{(\alpha,j)}}

\newcommand{\Y}{\Upsilon}
\newcommand{\anum}{\alpha=1,2,3,\quad j=1,2,...,n_{\alpha}}
\newcommand{\bnum}{\beta=1,2,3,\quad j=1,2,...,n_{\beta}}

\newcommand{\Om}{{\Omega}}
\newcommand{\omo}{\omega_0}
\newcommand{\Omt}{\Omega^\dagger}
\newcommand{\omos}{\omega^{*}_0}
\newcommand{\Phis}{\Phi^{*}}
\newcommand{\Psis}{\Psi^{*}}
\newcommand{\Rt}{{{\bf R}^3}}
\newcommand{\Rs}{{{\bf R}^6}}

\newcommand{\bC}{{\bf C}}
\newcommand{\Ct}{{{\bf C}^3}}
\newcommand{\Othree}{{\cal O}({\bf C}^3)}

\newcommand{\Osix}{{\cal O}({\bf C}^6)}

\newcommand{\Pilh}{\Pi_l^{(\rm hol)}}

\newcommand{\Sum}{\displaystyle\sum\limits}
\newcommand{\Int}{\displaystyle\int\limits}
\newcommand{\Min}[1]{\mathop{{\rm min}}\limits_{#1}}
\newcommand{\Max}[1]{\mathop{{\rm max}}\limits_{#1}}

\newcommand{\opla}{\mathop{\oplus}\limits}
\newcommand{\Times}{\mathop{\mbox{\large$\times$}}\limits}
\newcommand{\Bigcup}{\mathop{\bigcup}\limits}

\newcommand{\reduction}[2]{\left.\phantom{\bigl|} #1 \right|_{#2}}

\newcommand{\diag}{\mathop{\rm diag}}
\newcommand{\Img}{\mathop{\rm Im}}
\newcommand{\Real}{\mathop{\rm Re}}
\newcommand{\be}{\begin{equation}}
\newcommand{\ee}{\end{equation}}

\newcommand{\Frac}[2]{{\displaystyle\frac{#1}{#2}}}
\newtheorem{theorem}{\sc Theorem}
\newtheorem{lemma}{\sc Lemma}

\newtheorem{note}{\sc Remark}
\begin{document}
\thispagestyle{empty}
\large
\begin{center}
{\bf Bogoliubov Laboratory of Theoretical Physics\\[1.5mm] }
{\bf JOINT INSTITUTE FOR NUCLEAR RESEARCH\\[1.5mm]  }
{\bf 141980 Dubna (Moscow region), Russia}
\end{center} 
\vskip-3mm
\hrule{\hfill} 
\vskip 4.5cm
\normalsize

\hfill {\large Preprint JINR E5--95--46} 

\hfill LANL E-print {\tt nucl-th/9505029}
\bigskip

\bigskip

{\large
\begin{center}
          REPRESENTATIONS FOR THREE-BODY $T$--MATRIX \\[1.5mm]       
              ON UNPHYSICAL SHEETS:\,\, PROOFS%
\footnote{
\normalsize Published in Russian 
          in {\em Teoreticheskaya i Matematicheskaya 
          Fizika} {\bf 107} (1996) 478--500 [English translation 
          in {\em Theor. Math. Phys.}] under the title 
          ``Representations for three-body T-matrix 
          on unphysical sheets. II''}%
\footnote{
\normalsize The work supported in part by Academy of Natural Sciences 
             of RAS and International Science Foundation
             (Grant~\#~RFB000) }
\end{center}
}
\medskip

{\large
\centerline{  A.K.Motovilov\footnote{
\normalsize  E--mail: MOTOVILV@THSUN1.JINR.DUBNA.SU} }
}
\bigskip

\bigskip

\bigskip

\normalsize
\centerline{\bf Abstract}
\bigskip

A proof is given for the explicit representations which have been 
formulated in the author's previous work for the Faddeev components 
of three-body $T$--mat\-rix continued analytically on unphysical sheets 
of the energy Riemann surface. Also, the analogous representations for 
analytical continuation of the three-body scattering matrices and resolvent 
are proved. An algorithm to search for the three-body resonances on the 
base of the Faddeev differential equations is discussed.

\newpage
\normalsize
\setcounter{footnote}1
\setcounter{page}1
\section{\hspace*{-1em}. INTRODUCTION}
\label{SIntro}

The paper is a continuation of the author's work
~\cite{M1} devoted to studying a structure of the $T$--matrix, 
scattering matrices and resolvent of three--body Hamiltonian 
continued analytically on unphysical sheets of the energy Riemann surface. 

A central result of the paper~\cite{M1} consists in construction 
of the explicit representations for the continuation of three--body 
$T$--matrix on unphysical sheets in terms of this matrix itself 
taken on the physical one, as well as the scattering matrices. 
There were outlined only schemes to prove the representations above 
in Ref.~\cite{M1}. Main goal of the present work is to present 
a full proof. With the representations for $T$--matrix we base also 
analogous representations for analytical continuation of the scattering 
matrices and resolvent (see Ref.~\cite{M1}).

As in~\cite{M1} we suppose that interaction potentials are pairwise ones 
which decrease in the coordinate space not slower than exponentially. 
All the analysis is carried out on the base of the momentum 
space Faddeev integral equations~\cite{Faddeev63}, \cite{MF} 
for  components of the $T$--matrix. At that we find analytical 
continuation of the Faddeev equations as on neighboring unphysical sheets 
as on remote ones belonging to a certain part of the total three-body 
Riemann surface. A full description of the part under consideration 
see in Ref.~\cite{M1}.  The representations for the components of 
$T$--matrix on unphysical sheets arise as a result  of explicit 
solving the Faddeev equations continued in terms of the physical sheet. 

Note that a continuation of the  s--wave Faddeev equations on unphysical 
sheets neighboring with physical one, was made previously in 
the work~\cite{OrlovTur} (see also Ref.~\cite{Orlov})
in the case of separable pair potentials.

In the paper, we discuss also a practical meaning of the representations 
obtained. According to the representations [see Eqs.~(\ref{Ml3fin}),
(\ref{Slfin}) and (\ref{R3l})], the nontrivial singularities 
of the $T$--matrix as well as the scattering matrices and resolvent are 
determined, after the continuation of them on unphysical sheets, 
by singularities of the operators inverse to truncated 
scattering matrices on the physical sheet. 
Thus, the three--body resonances (i.e. the poles of the resolvent 
as well the $T$-- and 
scattering matrices) are actually those values of energy 
for which the scattering matrices, truncated 
in accordance with the index (number) of the unphysical sheet under 
consideration, have zero eigenvalue.  These properties 
of three--body scattering matrices are quite analogous to the familiar 
properties of the scattering matrices in problems of two particles 
and multichannel scattering problems with binary channels 
(see e.g., Refs.~\cite{Newton}--\cite{BohmQM} or~\cite{Orlov}, 
\cite{MotTMF}, \cite{MotYaF}).  For computations of three--body 
resonances as zeros of the truncated scattering matrices above, 
one can apply in particular, the differential formulation 
of the scattering problem~\cite{MF}, \cite{EChAYa} going 
on the complex plane of energy (physical sheet). 

The paper is organized as follows. 

In Sec.~\ref{SNotations2} we remember main notations of Ref.~\cite{M1}. 
The analytical continuation of the Faddeev equations on unphysical sheets 
is carried out in Sec.~\ref{SKernels}. Sec.~\ref{STRepres}
is devoted to deriving the explicit representations 
for the Faddeev components of the three--body $T$--matrix continued on 
unphysical sheets. 
The analogous representations are constructed in Sec.~\ref{SSmxl}  
for the scattering matrices and in Sec.~\ref{SResolvl}, for the resolvent.
In Sec.~\ref{SNumerMethod} we formulate an algorithm 
to calculate the three--body resonances on the base of the Faddeev 
differential equations in configuration space. 
\section{\hspace*{-1em}. NOTATIONS}
\label{SNotations2}
Throughout the paper we follow strictly by the conventions and 
notations adopted in Ref.~\cite{M1}.  Therefore we restrict ourselves 
here only to presenting for them a brief summary. Note at once
that at using formulae of the paper~\cite{M1} (it will take 
place rather often) we supply their number in Ref.~\cite{M1}
with the reference~``\cite{M1}''.

For the description of the system of three particles concerned 
in the momentum representation, we use the standard 
sets of reduced relative momenta~(\cite{M1}.2.1)  
$k\ad$, $p\ad$, \,\, $\an=1,2,3$,
which are usually combined into six-vectors 
$P= \lbrace k\ad, p\ad \rbrace$.
Transition from the pair $\{k\ad$, $p\ad\}$ to another 
one, $\{k\bd$, $p\bd\}$, corresponds to the rotation 
transform in ${\bf R}^{6}$,
$ k\ad=c\abd k\bd+s\abd p\bd$,\,\,
$ p\ad=-s\abd k\bd +c\abd p\bd$\,\,
with coefficients $c\abd$, $s\abd$~\cite{MF} depending 
on the particle masses only.

The Hamiltonian $H$ of the system is given by 
$(Hf)(P)=P^{2}f(P)+\sum_{\an=1}^{3} (v\ad f)(P),$
$P^2=k\ad^2+p\ad^2$, $f\in \cH_0\equiv L_2(\Rs)$,
where  $v\ad$, $\an=1,2,3,$ are pair potentials assumed for 
the sake of definiteness, to be local. This means that the kernel of 
each  $v\ad$ depends only on the difference of variables  
$k\ad$ and $k'\ad$,
$v\ad(k\ad, k'\ad)$ $=v\ad(k\ad - k'\ad)$.

We deal with two variants of the potentials $v\ad$.
In the first one, $v\ad (k)$ are holomorphic functions of variable 
$k\in\Ct$ satisfying the estimate~(\cite{M1}.2.2).
In the second variant, the potentials $v\ad (k)$ are holomorphic in 
$k$ in the strip $W_{2b} =\{ k: \, k\in\Ct,\,
|{\rm Im} k|<2b \}$ only and obey at $k\in W_{2b}$
the estimate~(\cite{M1}.2.3).
In the both variants $v\ad (-k)=$ $\overline{v\ad (k)}$, and this  
guarantees self--adjointness of the Hamiltonian $H$.

In the paper, the exposition is given for example of the second variant 
of potentials. Respective statements for the first one 
may be obtained from the statements of this work if to put in them, 
$b=+\infty$.  

By $h\ad$ we denote the Hamiltonians of the pair subsystems $\an$, 
$\an=1,2,3$.
Eigenvalues  $\lambda\ajd\in\sigma_{d}(h\ad)$ of $h\ad$, 
$\lambda\ajd < 0$, $ j = 1,2,...,n\ad$, $n\ad<\infty$, 
are enumerated taking into account their multiplicity: 
number of times to meet an eigenvalue in the numeration equals 
to its multiplicity. Maximal of these numbers is denoted by 
$\lambda_{\rm max}$,\,
$\lambda_{\rm max}=\Max{\an,j}\lambda\ajd<0.$
The notation $\psi\ajd(k\ad)$ is used for respective 
eigenfunctions.

We understand by $\sigma_d(H)$ and $\sigma_c(H)$  respectively 
the discrete and continuous components of the spectrum 
$\sigma(H)$ of the Hamiltonian 
$H$. Note that   $\sigma_c(H)=(\lambda_{\rm min},+\infty)$ with
$\lambda_{\rm min}= \Min{\an ,j} \lambda\ajd$.

The notation $H_0$ is adopted for the kinetic energy operator, 
$(H_0 f)(P)=$ $P^{2}f(P)$. By $R_0 (z)$ and $R(z)$ we denote 
the resolvents of $H_0$ and $H$, respectively: 
$R_0 (z)=(H_0-zI)^{-1}$ and $R(z)=(H -zI)^{-1}$ with $I$, the 
identity operator in $\cH_0$.

Let $ M\abd (z) = \delta\abd v\ad -v\ad R(z) v\bd$, $\an, \bn =1,2,3,$
be the components~\cite{Faddeev63}, \cite{MF} of 
the $T$--matrix $ T(z)=V-VR(z)V $ where $V=v_1 +v_1 +v_3 $.
The Faddeev equations~\cite{Faddeev63}, \cite{MF} for operators $M\abd$
read in matrix form as 
\be
\label{MFE}
M(z)=\bt (z) - \bt (z) \bRo (z) \Y M(z)
\ee
where  $\bRo (z)=\diag\{R_0(z),R_0(z),R_0(z)\}$ and by $\Y$
we understand the $3\!\times\! 3$--matrix with elements 
$\Y\abd=1-\delta\abd.$
Besides we use the notations 
$\bt (z)=\diag\{\bt_1(z),\bt_2(z),\bt_3(z)\}$. Here, the operators 
$\bt\ad(z)$, \,\, $\an=1,2,3,$ have the kernels 
$
{\bt}\ad(P,P',z)=t\ad(k\ad,k'\ad,z-p\ad^2)
\delta(p\ad-p'\ad)
$
where $t\ad(k,k',z)$ stand for the respective pair $T$--matrices $t\ad(z)$.
These $M$, $\bt$, $\bRo$
and $\Y$ are considered as operators in the Hilbert space 
${\cal G}_0=\opla_{\an=1}^{3} L_2(\Rs)$.

The resolvent $R(z)$ of $H$ is expressed by the matrix $M(z)$
as~\cite{MF}
\be
\label{RMR}
R(z)=R_0(z)-R_0(z) \Om M(z) \Omt R_0(z)
\ee
where $\Om$, $\Om: \cG_0\rightarrow\cH_0$, denotes operator 
defined as the matrix--row $\Om=(1,\,\, 1,\,\, 1)$. At the same time  
$\Omt=\Om^{*}=(1,\,\, 1,\,\, 1)^{\dagger}$.

Everywhere by $\sqrt{z-\lambda}$,\,\,\, $z\!\in\!\bC$,\,\,
$\lambda\!\in\!\bR,$\, we understand the main branch 
of the function 
$(z-\lambda)^{1/2}$.  Usually, by $\hat{q}$ we denote 
the unit vector in the direction  $q\!\in\!\bR^N$,\,\,
$\hat{q}={q}/{|q|}$, and by $S^{N-1}$ the unit sphere 
in $\bR^N$,\,\, $\hat{q}\!\in\!  S^{N-1}$.  The inner product in 
$\bR^N$ is denoted by $(\,\cdot\, ,\,\cdot\,)$.  Notation  
$\langle\,\cdot\, ,\,\cdot\, \rangle $ is used for 
inner products in Hilbert spaces. 

Let $\cH^{\aju}=L_2(\Rt)$ and $\cH^{\au}=\opla_{j=1}^{n\ad}
\cH^{\aju}$.  Notation $\Psi\ad$ is used for operator acting from
${\cal H}^{\au}$ to ${\cal H}_0$ as 
$
(\Psi\ad f)(P)=\sum_{j=1}^{n\ad} \psi\ajd (k\ad)f_j(p\ad) 
$.  
By $\Psi$ we understand the matrix--diagonal operator combined of 
$\Psi\ad$ as $\Psi=\diag\{\Psi_1,\Psi_2,\Psi_3\}$, 
and acting from  ${\cal H}_1=\opla_{\an=1}^{3} {\cal H}\au$ 
to ${\cal G}_0$.

The operators $\Phi\ad$ and $\Phi$ are obtained of $\Psi\ad$ and $\Psi$ by 
the replacement of functions $\psi\ajd(k\ad)$  with form--factors 
$\phi\ajd(k\ad)=(v\ad\psi\ajd)(k\ad)$, $\an=1,2,3$, $j=1,2,...,n\ad$.

By $ {\cal O}({\bf C}^{N})$ we denote the Fourier transform 
of the space $ C_{0}^{\infty}(\bR^{N})$. 

The operator $\rJ\ajd(z)$, \,\, $\anum$, realizes the restriction 
of functions 
$f(p\ad)$, $p\ad\in\Rt,$ on the energy shell 
$\lambda\ajd+|p\ad|^2=E$ at $z=E\pm i0$,\,\, $E>\lambda\ajd$, 
and then if possible, continues them analytically into 
a domain of complex values of energy $z$.  On ${\cal O}(\Ct)$, this 
operator acts as 
$\bigl(\rJ\ajd(z)f\bigr)(\hat{p}\ad)=f(\sqrt{z-\lambda\ajd}\hat{p}\ad).$ 
Notation  $\rJt\ajd(z)$ is used for the operator  ``transposed'' 
with respect to $\rJ\ajd(z)$ (see. Ref.~\cite{M1}).

The operator  $\rJo(z)$ is defined on  $\Osix$ analogously to $\rJ\ajd(z)$
by 
$
(\rJo(z)f)(\hat{P})=f(\sqrt{z}\hat{P}).
$
The notation $\rJot(z)$ is used for respective 
``transposed'' operator~\cite{M1}.

The operators $\rJ\ajd$ and $\rJt\ajd$ are combined in the diagonal 
matrices $\rJ\au(z)=$ $\diag\{ \rJ_{\alpha,1}(z),$
$...,\rJ_{\alpha,n\ad}(z) \}$ and $\rJ^{(\alpha)\dagger}(z)=$ $
\diag\{ \rJt_{\alpha,1}(z),...,$ $\rJt_{\alpha,n\ad}(z) \}. $ 
In their turn, we construct of the latter,  
the operators 
$\rJ_1 (z)=$ $\diag\{\rJ^{(1)}(z),$ $\rJ^{(2)}(z),$ $\rJ^{(3)}(z) \}$ 
and  
$\rJt_1 (z)=$ $\diag\{ \rJ^{(1)\dagger}(z),$ $\rJ^{(2)\dagger}(z),$
$\rJ^{(3)\dagger}(z) \}.$ 
Besides the listed ones, we use in the work, 
the block--diagonal operator $3\!\times\! 3$--matrices 
$\bJo(z)=\diag\{ \rJo(z),\rJo(z),\rJo(z) \} $ and  $\bJot(z)=\diag\{
\rJot(z),\rJot(z),\rJot(z) \} $ as well as operators $\bJ(z)=\diag\{
\rJo(z),\rJ_1(z) \} $ and $\bJt(z)=\diag\{ \rJot(z),\rJt_1(z) \}. $

Along with $\cH_0$, \,\, $\cG_0$ and $\cH_1$ described above,  
we consider the Hilbert spaces $\hat{\cH}_0= L_2(S^5),$
\,\, $\hat{\cG}_0=\opla_{\an=1}^3 \hat{\cH}_0$ and 
$\hat{\cH}_1=\opla_{\an=1}^{3}\hat{\cH}\au$ where 
$\hat{\cH}\au\equiv\opla_{j=1}^{n\ad} \hat{\cH}\aju, $
$\hat{\cH}\aju= L_2(S^2)$.  The identity operators in  $\hat{\cH}_0$,
$\hat{\cG}_0$,  $\hat{\cH}_1$ and $\hat{\cH}_0\oplus\hat{\cH}_1$
are denoted by $\hIo$, $\hbIo$, $\hIi$ and $\hbI$, respectively.

The operator--valued function $\cT(z)$,\,\,
$\cT(z):\,\cH_0\oplus\cH_1\rightarrow\cH_0\oplus\cH_1$,
of the variable $z\in\bC\setminus\overline{\sigma(H)}$ is defined by 
\be
\label{T3body}
\cT(z)\equiv \left(
\begin{array} {cc}
\Omega M(z)\Omega^{\dagger}   &    \Omega M(z)\Y\Psi  \\
\Psi^{*}\Y M(z)\Omega^{\dagger} &   \Psi^{*}(\Y\bv +\Y M(z)\Y)\Psi
\end{array}\right)
\ee
with $\bv=\diag\{ v_1, v_2, v_3 \}.$
The {\em truncated} three--body scattering matrices are expressed by $\cT(z)$
as
\be
\label{Slcut}
 S_l(z) \equiv \hat{\bf I} + (\tL\hat{\cT} L A)(z) \, \mbox{ and }\,
\St_l(z) \equiv \hat{\bf I} + (A L\hat{\cT}\tL)(z)
\ee
where  $\hat{\cT}(z)=(\bJ \cT \bJt)(z)$, \,\,
$\hat{\cT}(z):$
$\hat{\cH}_0 \oplus \hat{\cH}_1\rightarrow
\hat{\cH}_0 \oplus \hat{\cH}_1$.  The multi--index 
\be
\label{lmulti}
l=(l_0,l_{1,1},...,l_{1,n_1},l_{2,1},...,l_{2,n_2},l_{3,1},...,l_{3,n_3})
\ee
has the components $l_0=0$ or $l_0=\pm 1$
and $l\ajd=0$ or $l\ajd=1$, $\anum$.
Notations $L$ and $\tL$ are used for the diagonal matrices corresponding 
to the multi--index $l$: \,\,
$L=\diag\{ L_0,\, L_1\}$, \,\,\,
$\tL=\diag\{ |L_0|,L_1\}$, where $L_0=l_0$ and
$L_1=\diag\{l_{1,1},...,$ $l_{1,n_1},l_{2,1},...,$
$l_{2,n_2},l_{3,1},...,$ $l_{3,n_3}\}$.
By $A(z)$ we understand the diagonal matrix--function 
$A(z)=$ $\diag\{A_0(z),$ $A\ajd(z),$ $\anum\}$ with the elements  
$A_0(z)=-\pi i z^2$ and $A\ajd(z)=-\pi i\sqrt{z-\lambda\ajd}$. 

The notation $\Pi^{\rm (hol)}_{ll'}$ is used for the domain in variable  
$z\in\bC$ where $\bigl(L\hat{\cT}L'\bigr)(z)$ is a holomorphic 
operator--valued function. 
The matrices  $S_l(z)$ and $\St_l(z)$ as well as the products 
$\bigl( L_0 \bJo M \bigr)(z)$,\,\, $\bigl( L_1\rJ_1 \Psis\Y M\bigr)(z)$
and $\bigl( M \bJot L_0  \bigr)(z)$,\,\, $\bigl( M\Y\Psi\rJt_1 L_1 \bigr)(z)$
are holomorphic functions of $z$ on  domains 
$\Pilh=\Pi^{\rm (hol)}_{ll}$. A description of the domains 
$\Pi^{\rm (hol)}_{ll'}$ and $\Pilh$ see in Ref.~\cite{M1}, Sec~4.

We consider only a part of the total three--body Riemann 
surface. This part is denoted by $\Re$. 
Sheets $\Pi_l\subset\Re$ are generated by branching in the two--body, 
$z=\lambda\ajd$, $\anum$, and three--body, $z=0$, thresholds. 
When enumerating the sheets, the multi--index $l$ given by~(\ref{lmulti}) 
is used.
At $l_0=0$ its components $l\ajd$, $\anum$, can get arbitrary 
value among two numbers $0$ and $1$. In this case, $\Pi_l$ 
represents a copy of the complex plane $\bC$ cut along the ray 
$[\lambda_{\rm min},+\infty)$. If $l_0=\pm 1$ then the rest 
of components $l\ajd$, $\anum$, of $l$ are assumed be equal to 1.  
There is accepted that at $l_0=+1$ the sheet $\Pi_l$ coincides 
with the upper half--plane  
$\bC^{+}=\{z\in\bC:\, \Img z > 0\}$ and at $l_0=-1$, with the lower one, 
$\bC^{-}=\{z\in\bC:\, \Img z < 0\}$. We suppose additionally 
that the sheets $\Pi_l$ with $l_0=\pm 1$ are cut along the rays  
constituting together the set 
$Z_{\rm res}=\Bigcup_{\an=1}^3 Z_{\rm res}\au$.
Here, 
$
Z_{\rm res}\au=
\{z:\, z=z_r \rho, \, 1\leq\rho <+\infty,\,
z_r\in\sigma\au_{\rm res} \}
$
is a totality of the rays beginning in the resonance points 
$\sigma\au_{\rm res}$ of subsystem $\an$ and going to infinity 
along the directions 
$\hat{z}_r ={z_r}/{|z_{r}|}$, $z_r\in\sigma\au_{\rm res}$. 
A more detailed description of the surface $\Re$ and in particular,  
the way of sticking the sheets  $\Pi_l$ see in Sec.~5 of Ref.~\cite{M1}.

If all the  components of the multi--index $l$ are zero, $l_0=l\ajd=0,$ $\anum$, 
the sheet $\Pi_l$ is called the physical one, $\Pi_0$.    
The unphysical sheets $\Pi_l$ with $l_0=0$ are 
called the two--body ones since these sheets may be reached 
from $\Pi_0$ rounding the two--body thresholds 
$z=\lambda\ajd$ only, with no rounding the breakup threshold $z=0$.  
The sheets $\Pi_l$ at $l_0=\pm 1$ are called the three--body ones.
%
\section{\hspace*{-1em}. ANALYTICAL CONTINUATION OF FADDEEV EQUATIONS FOR 
\newline
                         COMPONENTS OF $T$--MATRIX ON UNPHYSICAL SHEETS }
\label{SKernels}
Goal of the present section consists in continuation on unphysical 
sheets of the surface $\Re$, of the absolute terms and kernels 
of the Faddeev equations (\ref{MFE}) and their iterations. The continuation 
is realized in a sense of generalized functions (distributions) over 
$\Osix$. Results of the continuation are represented in terms related 
with the physical sheet only. 

By $L\au$,\,\,\,  $L\au=L\au(l),$ we denote the diagonal matrices formed of 
the components $l_{\an,1},$ $l_{\an,2},...,$
$l_{\an,n\ad}$ of the multi--index $l$ of the sheet $\Pi_l\subset\Re$:\,\,\,
$L\au)=$ $\diag\{ l_{\an,1},$ $l_{\an,2},...,$ $l_{\an,n\ad} \}$.
At that $L_1(l)=\diag\{ L^{(1)},$ $L^{(2)},$ $L^{(3)} \}$ and
$L(l)=\diag\{L_0,$ $L_1\}$ с $L_0\equiv l_0$. Analogously,  
$A\au(z)=$ $\diag\{ A_{\an,1}(z),$ $A_{\an,2},...,$ $A_{\an,n\ad}(z)\}$ 
and $A_1(z)=\diag\{ A^{(1)}(z),$ $A^{(2)}(z),$ $A^{(3)}(z)\}$. Thus 
$A(z)=\diag\{A_0(z),A_1(z)\}$. 

By $\bs_{\an,l}(z)$ we understand the operator defined in $\hat{{\cal H}}_0$
as
\be
\label{spair6}
\bs_{\an,l}(z)=\hat{I}_0+\rJo(z)\bt\ad(z)\rJot(z)A_0(z)L_0,
\quad z\in\Pi_0.
\ee
It follows from Eq.~(\ref{spair6}) that $\bs_{\an,l}=\hat{I}_0$ at $l_0=0$.
If $l_0=\pm 1$ then according to Eqs.~(\cite{M1}.4.42)--\cite{M1}.4.44), 
the operator $\bs_{\an,l}(z)$ is defined for $z\in\cP_b\bigcap\bC^\pm$,
$\cP_b=\left\{ z:\,\,\,
\Real z>-b^2+\left.(\Img z)^2\right/(4b^{2}) \right\}$,
and acts on  $f\in \hat{\cH}_0$ as
\be
\label{spairh}
(\bs\adl(z)f)(\hP)=
\Int_{S^2} d\hk' s\ad(\hk\ad,\hk'\ad,z\cos^2\omega)
f(\cos\omega\ad\hk\ad',\sin\omega\ad\hp\ad)
\ee
where $\omega\ad,\hk\ad,\hp\ad$ stand for coordinates~\cite{MF} of the 
point $\hP$ on 
the hypersphere $S^5$, $\omega\ad\in\left[0,\pi/2\right]$,
$\hk\ad,\hp\ad \in S^2$.
For all this, 
$\hP=\lbrace \cos\omega\ad\hk\ad,\,\,\sin\omega\ad\hp\ad\rbrace$.
By $s\ad$ we denote the scattering matrix (\cite{M1}.2.16) for the pair 
subsystem $\alpha$.  As a matter of fact, $\bs\adl$ represents the scattering 
matrix $s\ad$ rewritten in the three--body momentum space. 

It follows immediately from Eq.~(\ref{spairh})
that if 
$z\in\cP_b\bigcap\bC^\pm\setminus {Z}\au_{\rm res}$ then there exists 
the bounded inverse operator $\bs\adl^{-1}(z)$,\,\,
\newline
$
(\bs\adl^{-1}(z)f)(\hP)=
\Int_{S^2} d\hk' s\ad^{-1}(\hk\ad,\hk'\ad,z\cos^2\omega\ad)
f(\cos\omega\ad\hk'\ad,\sin\omega\ad\hp\ad)
$
with $s\ad^{-1}(\hk,\hk',\zeta)$, 
the kernel of the inverse pair scattering matrix $s^{-1}\ad(\zeta)$.

The operator $\bs\adl^{-1}(z)$ becomes unbounded one 
at the boundary points $z$ situated on rims of the cuts 
(the ``resonance'' rays) included in $Z\au_{\rm res}$.

\begin{theorem}\label{ThKernels}\hspace*{-0.5em}{\sc .}
The absolute terms $\bt\ad(P,P',z)$
and kernels $(\bt\ad R_0)(P,P',z)$
of the Faddeev equations (\ref{MFE})
admit the analytical continuation in a sense of distributions over $\Osix$
both on two--body and three--body sheets  $\Pi_l$
of the Riemann surface  $\Re$.
The continuation on the sheet $\Pi_l$,
$l=(l_0,l_{1,1},...,$ $l_{1,n_1},l_{2,1},...,$
$l_{2,n_2},l_{3,1},...,l_{3,n_3})$, \,\,\,
 $l_0=0$, $l\bjd=0, 1,$
or $l_0=\pm 1$,  $l\bjd=1$
(in both cases $\bnum$), 
is written as 
\be
\label{tlRbig}
\bt^{l}\ad(z)\equiv \reduction{\bt\ad(z)}{\Pi_l}=
\bt\ad-L_0 A_0\bt\ad\rJot\bs\adl^{-1}\rJo\bt\ad-
\Phi\ad\rJ^{(\an)t}L\au A\au \rJ\au \Phi\ad^{*},
\ee
\be
\label{tr0l}
\reduction{[\bt\ad(z)R_0(z)]}{\Pi_l}=
\bt\ad^l (z)R_0^l(z)
\ee
where 
$
R_0^l(z)\equiv \reduction{R_0(z)}{\Pi_l}=
R_0(z)+L_0 A_0(z)\rJot(z)\rJo(z)
$
is the continuation~\cite{MotTMF} on $\Pi_l$
of the free Green function $R_0(z)$. If $l_0=0$ (and hence $\Pi_l$ is 
a two--body unphysical sheet),
the continuation in the form (\ref{tlRbig}), (\ref{tr0l})
is possible on the whole sheet $\Pi_l$. At $l_0=\pm 1$ (i.e. in the case 
when $\Pi_l$ is a three--body sheet)
the form (\ref{tlRbig}), (\ref{tr0l}) continuation is possible 
on the domain $\cP_b\bigcap\Pi_l$.
All the kernels in the right--hand parts of Eqs.~(\ref{tlRbig}) are 
taken on the physical sheet.
\end{theorem}
{\sc Proof} of the theorem we give for example of the most intricate 
continuation on three--body unphysical sheets 
$\Pi_l$ with  $l_0=\pm 1$.
For the sake of definiteness we consider 
the case $l_0=+1$.
For $l_0=-1$, the proof is quite analogous.

\begin{figure}
\centering
\unitlength=0.70mm
\special{em:linewidth .75pt}
\linethickness{.75pt}
\begin{picture}(159.33,95.54)
\emline{0.00}{7.67}{1}{69.67}{7.67}{2}
\emline{2.01}{24.85}{3}{3.66}{26.50}{4}
\emline{1.96}{26.50}{5}{3.72}{24.83}{6}
\emline{29.68}{24.85}{7}{31.33}{26.50}{8}
\emline{29.63}{26.50}{9}{31.39}{24.83}{10}
\emline{40.01}{24.85}{11}{41.66}{26.50}{12}
\emline{39.96}{26.50}{13}{41.72}{24.83}{14}
\emline{36.34}{33.18}{15}{37.99}{34.83}{16}
\emline{36.30}{34.83}{17}{38.06}{33.16}{18}
\emline{16.31}{24.87}{19}{17.96}{26.51}{20}
\emline{16.26}{26.51}{21}{18.02}{24.84}{22}
\bezier{372}(55.67,25.67)(83.33,89.33)(94.33,68.00)
\bezier{132}(94.33,68.33)(100.33,57.00)(85.00,43.67)
\bezier{372}(55.67,25.67)(83.33,89.33)(94.33,68.00)
\bezier{372}(55.67,25.67)(83.33,89.33)(94.33,68.00)
\bezier{132}(94.33,68.33)(100.33,57.00)(85.00,43.67)
\emline{83.67}{25.33}{23}{159.33}{25.33}{24}
\emline{159.33}{25.33}{25}{159.00}{25.33}{26}
\emline{82.34}{31.18}{27}{83.99}{32.83}{28}
\emline{82.30}{32.83}{29}{84.06}{31.16}{30}
\emline{67.01}{61.52}{31}{68.66}{63.17}{32}
\emline{66.96}{63.17}{33}{68.72}{61.50}{34}
\emline{37.01}{61.52}{35}{38.66}{63.17}{36}
\emline{36.96}{63.16}{37}{38.72}{61.50}{38}
\emline{0.00}{53.33}{39}{37.00}{34.00}{40}
\emline{23.26}{95.21}{41}{37.94}{62.19}{42}
\emline{37.94}{62.19}{43}{37.94}{62.28}{44}
\emline{159.00}{51.00}{45}{82.97}{31.95}{46}
\bezier{228}(85.04,43.54)(59.29,24.05)(83.47,25.33)
\put(69.55,7.76){\circle*{3.00}}
\put(55.67,25.67){\circle*{3.00}}
\put(56.00,20.00){\makebox(0,0)[cc]{0}}
\put(69.67,2.00){\makebox(0,0)[cc]{$z$}}
\put(115.67,19.66){\makebox(0,0)[cc]{(\,$\rho$\,)}}
\put(18.67,1.66){\makebox(0,0)[cc]{($\lambda$)}}
\put(18.33,19.00){\makebox(0,0)[cc]{$\lambda_{\alpha,j}$}}
\emline{123.65}{26.14}{47}{123.65}{24.38}{48}
\emline{123.65}{24.38}{49}{127.44}{25.40}{50}
\emline{127.44}{25.40}{51}{123.51}{26.14}{52}
\emline{123.65}{25.74}{53}{127.17}{25.40}{54}
\emline{127.17}{25.40}{55}{127.10}{25.40}{56}
\emline{127.10}{25.40}{57}{123.58}{24.86}{58}
\emline{123.58}{24.86}{59}{123.65}{25.06}{60}
\emline{123.65}{25.06}{61}{126.76}{25.33}{62}
\emline{126.76}{25.33}{63}{123.58}{25.53}{64}
\emline{22.32}{8.48}{65}{22.32}{6.72}{66}
\emline{22.32}{6.72}{67}{26.10}{7.73}{68}
\emline{26.10}{7.73}{69}{22.18}{8.48}{70}
\emline{22.32}{8.07}{71}{25.83}{7.73}{72}
\emline{25.83}{7.73}{73}{25.77}{7.73}{74}
\emline{25.77}{7.73}{75}{22.25}{7.19}{76}
\emline{22.25}{7.19}{77}{22.32}{7.39}{78}
\emline{22.32}{7.39}{79}{25.43}{7.66}{80}
\emline{25.43}{7.66}{81}{22.25}{7.87}{82}
\emline{158.00}{95.33}{83}{158.00}{95.33}{84}
\emline{158.00}{95.33}{85}{158.00}{95.33}{86}
\emline{78.18}{95.36}{87}{78.18}{95.36}{88}
\emline{67.82}{62.40}{89}{78.52}{95.54}{90}
\end{picture}
\caption{ 
Deformation of the integration contour over variable $\rho$. 
The integration contours over $\rho$ and $\lambda$ are denoted by  
letters in brackets. The cross ``$\times$'' denotes the eigenvalues  
$\lambda_{\alpha,j}$ of $h_\alpha$ on the negative half--axis
of the physical sheet and the pair resonances belonging to the set  
$\sigma_{\rm res}^{(\alpha)}$ 
on the sheet $\Pi_l$,\,\, $l_0=+1$. Also, there are denoted the cuts 
on $\Pi_l$,\,\, $l_0=+1,$ 
beginning at the points of $\sigma_{\rm res}^{(\alpha)}$. 
}
\label{figContour-mu}
\end{figure}
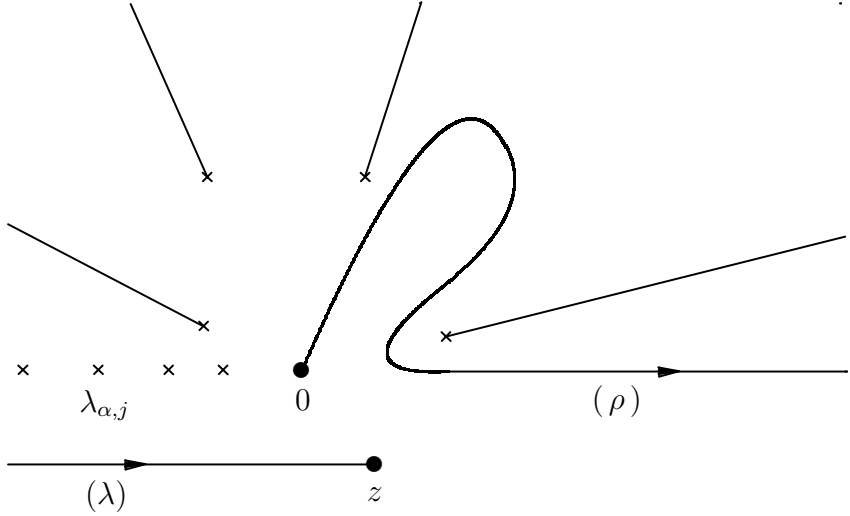

Let us consider at $z\in\Pi_0$, $\Img z<0$ the bilinear form
\be
\label{ftr0f}
(f,\bt\ad R_0(z)f')=
\Int_{\Rt}dk\Int_{\Rt}dk'\Int_{\Rt}dp\,
\Frac{t\ad(k,k',z-p^2)}{k'^2+p^2-z}\,
\tilde{f}(k,k',p)
\ee
with $\tilde{f}(k,k',p)=f(k,p)f'(k',p)$,
$f,f'\in\Osix$,
$k=k\ad,$ $k'=k\ad'$, $p=p\ad$.
Making replacements of variables $|k'|\rightarrow\rho=|k'|^2$,
$|p|\rightarrow \lambda=z-|p|^2$ we find that the integral 
(\ref{ftr0f}) turns into 
\be
\label{ftr0f1}
\Frac{1}{4}\Int_\Rt dk\Int_{S^2}d\hk'\Int_{S^2}d\hp
\Int_{z-\infty}^{z}d\lambda\sqrt{z-\lambda}\Int_0^\infty
d\rho\sqrt{\rho}\,\Frac{t\ad(k,\sqrt{\rho}\hk',\lambda)}{\rho-\lambda}\,
\tilde{f}(k,\sqrt{\rho}\hk',\sqrt{z-\lambda}\hp).
\ee
Existence of the analytical continuation 
of the kernel $(\bt\ad R_0)(z)$ on the sheet 
$\Pi_l$, $l_0=\pm 1$, follows from a possibility to deform continuously 
the contour of integration over variable $\rho$
to arbitrary sector of the analyticity domain 
$\cP_b\bigcap\sigma\au_{\rm res}$
of the integrand in variable $\lambda$ in the way demonstrated 
in Fig.~\ref{figContour-mu}.
Besides, this is connected with a possibility 
at moving of $z$ from $\Pi_0$ to $\Pi_l$, $l_0=+1$,
to make a necessary deformation 
of the integration path in variable $\lambda$ in such a way 
that this path is separated from the integration contour in 
variable $\rho$.

\begin{figure}
\centering
\unitlength=0.70mm
\special{em:linewidth .75pt}
\linethickness{.75pt}
\begin{picture}(159.33,95.54)
\emline{2.01}{24.85}{1}{3.66}{26.50}{2}
\emline{1.96}{26.50}{3}{3.72}{24.83}{4}
\emline{29.68}{24.85}{5}{31.33}{26.50}{6}
\emline{29.63}{26.50}{7}{31.39}{24.83}{8}
\emline{40.01}{24.85}{9}{41.66}{26.50}{10}
\emline{39.96}{26.50}{11}{41.72}{24.83}{12}
\emline{36.34}{33.18}{13}{37.99}{34.83}{14}
\emline{36.30}{34.83}{15}{38.06}{33.16}{16}
\emline{16.31}{24.87}{17}{17.96}{26.51}{18}
\emline{16.26}{26.51}{19}{18.02}{24.84}{20}
\emline{83.67}{25.33}{21}{159.33}{25.33}{22}
\emline{159.33}{25.33}{23}{159.00}{25.33}{24}
\emline{82.34}{31.18}{25}{83.99}{32.83}{26}
\emline{82.30}{32.83}{27}{84.06}{31.16}{28}
\emline{67.01}{61.52}{29}{68.66}{63.17}{30}
\emline{66.96}{63.17}{31}{68.72}{61.50}{32}
\emline{37.01}{61.52}{33}{38.66}{63.17}{34}
\emline{36.96}{63.16}{35}{38.72}{61.50}{36}
\emline{0.00}{53.33}{37}{37.00}{34.00}{38}
\emline{23.26}{95.21}{39}{37.94}{62.19}{40}
\emline{37.94}{62.19}{41}{37.94}{62.28}{42}
\emline{159.00}{51.00}{43}{82.97}{31.95}{44}
\put(55.67,25.67){\circle*{3.00}}
\put(57.67,17.67){\makebox(0,0)[cc]{0}}
\put(121.67,18.66){\makebox(0,0)[cc]{$\Gamma_2$}}
\put(22.60,13.90){\makebox(0,0)[cc]{$\lambda_{\alpha,j}$}}
\emline{123.65}{26.14}{45}{123.65}{24.38}{46}
\emline{123.65}{24.38}{47}{127.44}{25.40}{48}
\emline{127.44}{25.40}{49}{123.51}{26.14}{50}
\emline{123.65}{25.74}{51}{127.17}{25.40}{52}
\emline{127.17}{25.40}{53}{127.10}{25.40}{54}
\emline{127.10}{25.40}{55}{123.58}{24.86}{56}
\emline{123.58}{24.86}{57}{123.65}{25.06}{58}
\emline{123.65}{25.06}{59}{126.76}{25.33}{60}
\emline{126.76}{25.33}{61}{123.58}{25.53}{62}
\emline{158.00}{95.33}{63}{158.00}{95.33}{64}
\emline{158.00}{95.33}{65}{158.00}{95.33}{66}
\emline{78.18}{95.36}{67}{78.18}{95.36}{68}
\emline{67.82}{62.40}{69}{78.52}{95.54}{70}
\bezier{348}(55.43,25.51)(82.83,88.75)(97.70,78.10)
\bezier{376}(55.59,25.19)(115.70,61.20)(98.01,77.95)
\emline{55.28}{25.35}{71}{83.45}{25.35}{72}
\emline{55.75}{25.35}{73}{90.03}{70.43}{74}
\put(89.79,70.60){\circle*{3.00}}
\emline{55.59}{70.59}{75}{55.59}{25.35}{76}
\emline{55.67}{70.33}{77}{0.00}{70.33}{78}
\emline{91.07}{80.19}{79}{91.07}{78.43}{80}
\emline{91.07}{78.43}{81}{94.86}{79.45}{82}
\emline{94.86}{79.45}{83}{90.93}{80.19}{84}
\emline{91.07}{79.79}{85}{94.59}{79.45}{86}
\emline{94.59}{79.45}{87}{94.52}{79.45}{88}
\emline{94.52}{79.45}{89}{91.00}{78.91}{90}
\emline{91.00}{78.91}{91}{91.07}{79.11}{92}
\emline{91.07}{79.11}{93}{94.18}{79.38}{94}
\emline{94.18}{79.38}{95}{91.00}{79.58}{96}
\emline{91.10}{78.65}{97}{94.77}{79.46}{98}
\emline{94.77}{79.46}{99}{91.04}{80.01}{100}
\emline{12.73}{71.19}{101}{12.73}{69.43}{102}
\emline{12.73}{69.43}{103}{16.52}{70.45}{104}
\emline{16.52}{70.45}{105}{12.59}{71.19}{106}
\emline{12.73}{70.79}{107}{16.25}{70.45}{108}
\emline{16.25}{70.45}{109}{16.18}{70.45}{110}
\emline{16.18}{70.45}{111}{12.66}{69.91}{112}
\emline{12.66}{69.91}{113}{12.73}{70.11}{114}
\emline{12.73}{70.11}{115}{15.84}{70.38}{116}
\emline{15.84}{70.38}{117}{12.66}{70.58}{118}
\emline{12.77}{69.65}{119}{16.44}{70.46}{120}
\emline{16.44}{70.46}{121}{12.71}{71.01}{122}
\put(40.78,25.42){\circle{8.75}}
\put(30.29,25.42){\circle{8.75}}
\put(17.17,25.42){\circle{8.75}}
\put(2.74,25.42){\circle{8.75}}
\emline{2.01}{22.02}{123}{2.01}{20.42}{124}
\emline{2.01}{20.42}{125}{4.51}{21.36}{126}
\emline{4.51}{21.36}{127}{2.01}{21.98}{128}
\emline{1.97}{21.71}{129}{4.51}{21.32}{130}
\emline{4.51}{21.32}{131}{2.01}{20.69}{132}
\emline{2.01}{20.97}{133}{4.51}{21.28}{134}
\emline{1.97}{21.40}{135}{4.54}{21.32}{136}
\emline{16.44}{22.31}{137}{16.44}{20.71}{138}
\emline{16.44}{20.71}{139}{18.93}{21.65}{140}
\emline{18.93}{21.65}{141}{16.44}{22.27}{142}
\emline{16.40}{22.00}{143}{18.93}{21.61}{144}
\emline{18.93}{21.61}{145}{16.44}{20.98}{146}
\emline{16.44}{21.26}{147}{18.93}{21.57}{148}
\emline{16.40}{21.69}{149}{18.97}{21.61}{150}
\emline{29.55}{22.17}{151}{29.55}{20.57}{152}
\emline{29.55}{20.57}{153}{32.05}{21.50}{154}
\emline{32.05}{21.50}{155}{29.55}{22.13}{156}
\emline{29.51}{21.85}{157}{32.05}{21.46}{158}
\emline{32.05}{21.46}{159}{29.55}{20.84}{160}
\emline{29.55}{21.11}{161}{32.05}{21.42}{162}
\emline{29.51}{21.54}{163}{32.09}{21.46}{164}
\emline{40.05}{22.17}{165}{40.05}{20.57}{166}
\emline{40.05}{20.57}{167}{42.55}{21.50}{168}
\emline{42.55}{21.50}{169}{40.05}{22.13}{170}
\emline{40.01}{21.85}{171}{42.55}{21.46}{172}
\emline{42.55}{21.46}{173}{40.05}{20.84}{174}
\emline{40.05}{21.11}{175}{42.55}{21.42}{176}
\emline{40.01}{21.54}{177}{42.58}{21.46}{178}
\emline{71.84}{48.08}{179}{73.16}{47.05}{180}
\emline{73.16}{47.05}{181}{74.85}{50.59}{182}
\emline{74.85}{50.59}{183}{71.80}{48.04}{184}
\emline{72.05}{47.92}{185}{74.85}{50.55}{186}
\emline{74.85}{50.55}{187}{72.91}{47.18}{188}
\emline{72.71}{47.34}{189}{74.68}{50.43}{190}
\emline{72.25}{47.75}{191}{74.68}{50.35}{192}
\emline{-10.33}{70.33}{193}{1.33}{70.33}{194}
\emline{0.41}{53.05}{195}{-10.78}{58.73}{196}
\put(6.33,75.00){\makebox(0,0)[cc]{$G_2$}}
\put(84.67,53.33){\makebox(0,0)[cb]{$G_1$}}
\put(94.33,71.33){\makebox(0,0)[cc]{$z$}}
\put(104.67,80.33){\makebox(0,0)[cc]{$\Gamma_1$}}
\end{picture}
\caption{ 
The final location of the integration contours over variables
$\rho$\,\, $(\Gamma_1\bigcup\Gamma_2)$ and $\lambda$ \,\, $(G_1\bigcup G_2)$. 
The contour  $\Gamma_1$ represents a loop going clockwise around 
the contour $G_1$, the line segment 
$[0,\,\, z]$;\, $\Gamma_2=[0,\,+\infty)$; 
\, $G_2=(z-\infty,\, i\,{\rm Im}\, z]\bigcup[i\,{\rm Im}\, z,\, 0)$. 
}
\label{figContour-fin}
\end{figure}
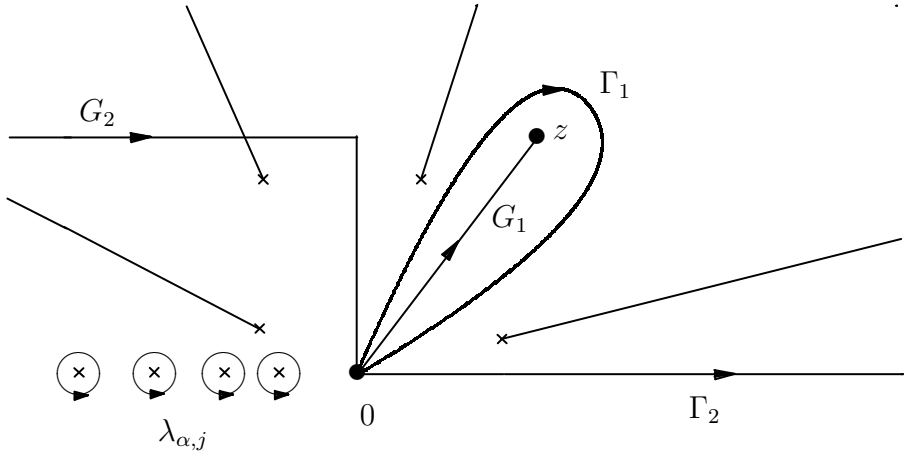

To obtain the representation (\ref{tr0l})
at a concrete point $z=z_0$, we choose a special final location 
of the integration contours in variables 
$\lambda$ and $\rho$ after consistent deforming them 
(see Fig.~\ref{figContour-fin}).
Singularity of inner integral (over variable $\rho$)
remains integrable after such deformation 
due to presence of the factor  $\sqrt{\rho}$.
As a whole the integral (\ref{ftr0f1}) turns into 
\be
\label{formpf}
\begin{array}{c}
 \Frac{1}{4}\Int_\Rt dk\Int_{S^2} d\hk'\Int_{S^2} d\hp\times      \\
 \times\left\{
\Int_{G_1} d\lambda\sqrt{z-\lambda} \Int_{\Gamma_1\bigcup\Gamma_2}
d\rho\sqrt{\rho}\,
\Frac{t'\ad(k,\sqrt{\rho}\hk',\lambda)}{\rho-\lambda}\,
\tilde{f}(k,\sqrt{\rho}\hk',\sqrt{z-\lambda}\hp)+  \right.\\
 +\Int_{G_2} d\lambda\sqrt{z-\lambda} \Int_{\Gamma_1\bigcup\Gamma_2}
d\rho\sqrt{\rho}\,
\Frac{t\ad(k,\sqrt{\rho}\hk',\lambda)}{\rho-\lambda}\,
\tilde{f}(k,\sqrt{\rho}\hk',\sqrt{z-\lambda}\hp)+  \\
 \left. +\Sum_{j=1}^{n\ad}2\pi i \sqrt{z-\lambda\ajd}
\Int_{0}^{+\infty}d\rho\sqrt{\rho}\,
\Frac{\phi\ajd(k)\overline{\phi}\ajd(k')}{\rho-\lambda\ajd}\,
\tilde{f}(k,\sqrt{\rho}\hk',\sqrt{z-\lambda\ajd}\hp)
\right\}
\end{array}
\ee
where $t'\ad$ denotes the pair $T$--matrix $t\ad(z)$ continued on 
the second sheet
(as regards $t\ad(\lambda)$, the contour 
$G_1\ni\lambda$ belongs to its second sheet). 
The last term arises as a result of taking residues in the points 
$\lambda\ajd\in\sigma_d(h\ad)$.

Evidently, the domain of variable $z\in\Pi_l$,\, \, $l_0=+1$,
where one can continue analytically the function (\ref{ftr0f}) in the 
form  (\ref{formpf}) to, is determined by the conditions  $\Gamma_1\subset\cP_b$ and
$\Gamma_1\bigcap Z\au_{\rm res}=\emptyset$. These conditions may be 
satisfied at $z\in\cP_b$ only.

Note that value of the inner integrals over  $\Gamma_1$
at $\lambda\in G_1$ are determined by residues at the points 
$\rho=\lambda$. At the same time 
$\int_{G_2}d\lambda...\int_{\Gamma_1}...=0$
since at $\lambda\in G_2$ the functions under the integration sign are 
holomorphic in $\rho\in{\rm Int\,}\Gamma_1$. 
Therefore 
\begin{eqnarray}
\lefteqn{\reduction{(f,\bt\ad R_0(z)f')}{z\in\Pi_l,\, l_0=+1} =
\Frac{1}{4}\Int_\Rt dk\Int_{S^2} d\hk'\Int_{S^2} d\hp\,\times }\nonumber  \\
 & & \times \left\{
\Int_{G_1} d\lambda\sqrt{z-\lambda} (-2\pi i) \sqrt{\lambda}\,
t'\ad(k,\sqrt{\lambda}\hk',\lambda)
\tilde{f}(k,\sqrt{\lambda}\hk',\sqrt{z-\lambda}\,\hp)+ \right. \nonumber \\
 & & +\Int_{G_1} d\lambda\sqrt{z-\lambda} \Int_{\Gamma_2}
d\rho\sqrt{\rho}\,
\Frac{t\ad(k,\sqrt{\rho}\hk',\lambda)+\pi i \sqrt{\lambda}\,
\tau\ad(k,\sqrt{\rho}\hk',\lambda)}{\rho-\lambda}
\tilde{f}(k,\sqrt{\rho}\hk',\sqrt{z-\lambda}\hp)+  \label{tR0fin}\\
 & & +\Int_{G_2} d\lambda\sqrt{z-\lambda} \Int_{\Gamma_2}
d\rho\sqrt{\rho}\,
\Frac{t\ad(k,\sqrt{\rho}\hk',\lambda)}{\rho-\lambda}
\tilde{f}(k,\sqrt{\rho}\hk',\sqrt{z-\lambda}\,\hp)+  \nonumber \\
 & & \left. +\Sum_{j=1}^{n\ad}2\pi i \sqrt{z-\lambda\ajd}
\Int_{0}^{+\infty}d\rho\sqrt{\rho}\,
\Frac{\phi\ajd(k)\overline{\phi}\ajd(k')}{\rho-\lambda\ajd}\,
\tilde{f}(k,\sqrt{\rho}\hk',\sqrt{z-\lambda\ajd}\,\hp)
\right\}. \nonumber
\end{eqnarray}
In the second summand of Eq.~(\ref{tR0fin}) we have used the 
representation (\cite{M1}.3.2) of for the pair 
$T$--matrix continued on the second sheet. Look at the expression for   
$\tau\ad(k,k',\zeta)$  in Ref.~\cite{M1}, Sec.~3. 
Remember that 
$t'\ad(\zeta)=t\ad(\zeta)+\pi i\sqrt{\zeta}\,\,\tau\ad(\zeta)$. 

Joining the summands including $t\ad$ on the physical sheet, in 
the alone  integral $\int_{G_1\bigcup G_2}$...
and using then  the holomorphness of the function under the integration sign 
in variable $\lambda$,  we 
straighten the contour $G_1\bigcup G_2$ turning it into the ray 
$(z-\infty,z]$. As a result we get the  bilinear form corresponding 
to the product $\bigl(\bt\ad R_0\bigr)(z)$ taken on the physical sheet. 

The last term of the expression~ (\ref{tR0fin})
corresponds to the kernel of 
$-\Phi\ad\rJ^{(\an)\dagger}L\au A\au\rJ\au \Phi^{*}R_0$.

Backing in the rest of summands including $t'\ad$
and $\tau\ad$ to the initial variables $k'$, $p'$  and utilizing then 
the definition (\ref{spair6}), we find that these summands 
correspond to the expression 
$$
L_0 A_0\left[\bt\ad -L_0 A_0 \rJot\bs\adl^{-1}\rJo\bt\ad\right]\rJot\rJo
- L_0 A_0 \bt\ad\rJot\bs\adl^{-1}\rJo\bt\ad R_0.
$$
Gathering the results obtained we reveal that the analytical 
continuation of $\bt\ad R_0$ on the sheet $\Pi_l,$ $l_0=+1$,
looks as 
\begin{eqnarray}
\lefteqn{\reduction{[\bt\ad R_0(z)]}{\Pi_l}
 =(\bt\ad-L_0 A_0\bt\ad\rJot\bs\adl^{-1}\rJo\bt\ad-
\Phi\ad\rJ^{(\an)\dagger}L\au A\au \rJ\au \Phi^{*}\ad)\times } \nonumber\\
\label{polufinal}
 & & \times (R_0+L_0 A_0\rJot\rJo) +
    L_0 A_0 \Phi\ad\rJ^{(\an)\dagger}
L\au A\au \rJ\au \Phi^{*}\ad\rJot\rJo.
\end{eqnarray}
To be convinced in the factorization (\ref{tr0l}), is sufficient to 
note that the last summand of (\ref{polufinal}) equals to zero. 
Indeed, one can check easily that at 
$\Img z\neq 0$ or $\Img z=0$ and $z>\Max{j} \lambda\ajd$ 
the following equalities take place 
\be
\label{JFJ}
(\rJ\au\Phi\ad^{*}\rJot)(z)=0,\quad  (\rJo\Phi\ad\rJ\aut)(z)=0.
\ee
Thereby the last term of~(\ref{polufinal}) disappears and hence, 
Eq.~(\ref{tr0l}) is true.
This completes the proof. 
\begin{note}\label{Nthetaz}\hspace*{-0.5em}{\sc .}
{\rm As a matter of fact, the kernel 
$\reduction{[\bt\ad R_0](z)}{\Pi_l} $
corresponds to the two--body problem and thereby it has to be 
translationally invariant with respect to variable  $p\ad$.
This fact may be understood if one 
introduces the generalized function (distribution) 
$\theta_z(p)$ over $\Othree$ acting as 
$
(\theta_z ,f)=\Frac{1}{2}\Int_{\gamma_z}
d\xi\sqrt{\xi}\Int_{S^2} d\hp f(\sqrt{\xi}\hp) 
$
where  $\gamma_z$ is the line segment connecting the points 
$\xi=0$ and $\xi=z$.
It follows from the representation  $(\ref{tR0fin})$ that the kernel of 
$\reduction{[\bt\ad R_0](z)}{\Pi_l}$  may be rewritten as 
\begin{eqnarray*}
\lefteqn{(\bt\ad R_0)^l(P,P',z)=\delta(p-p')\left\{
\Frac{t\ad(k,k',z-p^2)}{k'^2+p^2-z}+          \right.   } \\
 & & +\pi i L_0 \left[ \theta_z(p)\sqrt{z-p^2}\,\,
\Frac{\tau\ad(k,k',z-p^2)}{k'^2+p^2-z}-
\theta_z(k')\sqrt{z-k'^2}\,\, t'\ad(k,k',k'^2)
\Frac{\delta(\sqrt{z-k'^2}-|p|)}{|p|^2}\right] +  \\
 & & \left. +\Sum_{j=1}^{n\ad}2\pi i\,\, l\ajd\sqrt{z-\lambda\ajd}\,\,
\Frac{\phi\ajd(k)\overline{\phi}\ajd(k')}{k^2-\lambda\ajd}\,\cdot\,
\Frac{\delta(|p|-\sqrt{z-\lambda\ajd})}{|p|^2}
\right\},\\
 & & k=k\ad,\; k'=k'\ad,\; p=p\ad,\; p'=p'\ad,
\end{eqnarray*}
where due to the presence of the factor $\delta(p-p')$, the translation 
invariance is emphasized explicitly. Analogously 
\begin{eqnarray*}
\bt\ad^l(P,P',z) &=& \left\{
t\ad(k,k',z-p^2)+\pi i\, L_0\, \theta_z(p)
\sqrt{z-p^2}\,\,\tau\ad(k,k',z-p^2) +                \right. \\
    &+& \Sum_{j=1}^{n\ad}2\pi i\, l\ajd\sqrt{z-\lambda\ajd}\,
\left.
{ \phi\ajd(k)\overline{\phi}\ajd(k') }\,\cdot\,
\Frac{\delta(|p|-\sqrt{z-\lambda\ajd})}{|p|^2}
 \right\}
\delta(p-p').
\end{eqnarray*}
}
\end{note}

Using Eqs.~(\ref{tlRbig}) and (\ref{tr0l}) one can 
present the Faddeev equations (\ref{MFE}) continued 
on the sheet $\Pi_l$ in the matrix form
\be
\label{MFEl}
M^l(z)=\bt^l(z)-\bt^l(z)\bRo^l(z)\Y M^l(z)
\ee
where
\be
\label{tltot}
\bt^l(z)=\bt-
L_0 A_0 \bt\bJot\bs_l^{-1}\bJo\bt-\Phi\rJt_1 L_1 A_1\rJ_1\Phi^{*},
\ee
\be
\label{R0ltot}
\bRo^l(z)=\bRo(z)+L_0 A_0(z)\bJot(z)\bJo(z).
\ee
Here, 
 $\bs_l(z)=\diag\{\bs_{1,l}(z),\bs_{2,l}(z),\bs_{3,l}(z)\}$.
By $M^l(z)$ we understand the supposed analytical continuation 
on the sheet $\Pi_l$ of the matrix $M(z)$.
\begin{lemma}\label{LPossibility}\hspace*{-0.5em}{\sc .}
For each two--body unphysical sheet $\Pi_l$ of the surface 
$\Re$ there exists such a path from the physical sheet $\Pi_0$ 
to the domain $\Pi_l^{\rm (hol)}$ in $\Pi_l$ going only 
on two--body unphysical sheets $\Pi_{l'}$ that moving by this path,  
the parameter $z$ stays always in respective domains 
$\Pi_{l'}^{\rm (hol)}\subset\Pi_{l'}$.
\end{lemma}
{\sc Proof.} Let us use the principle of mathematical induction. 
To make this, at the beginning we arrange the branching points 
$\lambda\ajd,$ $\anum,$ in nondecreasing order redenoting them as
$\lambda_1,\lambda_2,...,\lambda_m$,
$m\leq\Sum_{\an} n\ad,$ $\lambda_1 <\lambda_2 <...< \lambda_m$,
and putting $\lambda_{m+1}=0.$
Let the multi--index $l=(l_1,l_2,...,l_m)$ correspond 
temporarily namely to this enumeration.
As previously, $l_j=0$
if the sheet $\Pi_l$ is related to the main branch of the function 
$(z-\lambda_j)^{1/2}$ else $l_j=1$. 
The index $l_0$ is omitted in these temporary notations. 

It is clear that the transition of z from the physical sheet $\Pi_0$
across the segment $(\lambda_1,\lambda_2)$ on the neighboring 
unphysical sheet $\Pi_{l^{(1)}}$ 
(into the domain $\Pi_{l^{(1)}}^{\rm (hol)}$), \, 
$l^{(1)}=(l_1^{(1)},l_2^{(1)},...,l_m^{(1)})$
with $l_1^{(1)}=1$ and $l_j^{(1)}=0$ at $j\neq 1,$
is possible by definition  of the domain 
$\Pi_{l^{(1)}}^{\rm (hol)}$ (see Ref.~\cite{M1}, Sec~4).
According to Lemmas~\cite{M1}.1 and~\cite{M1}.2,
if $z$ belongs to $\Pi_{l^{(1)}}^{\rm (hol)}$, it may be  
lead to the real axis in the interval 
$(\lambda^{(1)},+\infty)$
with certain $\lambda^{(1)} <\lambda_1. $ Remaining in 
 $\Pi_{l^{(1)}}^{\rm (hol)}$ the point $z$ may even go around 
the threshold 
$\lambda_1$ crossing the real axis in the segment 
$(\lambda^{(1)}, \lambda_1). $
Thus, the parameter $z$ may be lead from the sheet $\Pi_{l^{(1)}}$ 
on the each neighboring unphysical sheet 
and in particular, on the sheet $\Pi_l$
related to  $l_1=0,$    $l_2=1$,   $l_j=0,$  $j\geq 3.$
Transition of $z$ from $\Pi_0$
across the segment $(\lambda_2,\lambda_3)$
on the sheet $\Pi_l$
with $l_1=l_2=1,$    $l_j=0,$   $j\geq 3,$
is always possible.

We suppose further that the parameter $z$ may be carried in this manner
from $\Pi_0$  on all the two--body unphysical sheets 
$\Pi_{l^{(k)}}$ defined by the conditions $l_j^{(k)}=0,$   $j >k. $
It assumed also that during the carrying, $z$ always remains 
in the domains $\Pi_{l^{(k)}}^{\rm (hol)}$ of these sheets 
and does not visit different sheets.   
It follows from Lemmas~\cite{M1}.1 and~\cite{M1}.2 that if $z$ stays 
in the domain $\Pi_{l^{(k)}}^{\rm (hol)}$ of each sheet of the type 
described then wittingly, it can be lead to the real axis in the segment 
$(\lambda^{(k)},+\infty)$ with certain $\lambda^{(k)} < \lambda_k.$
Hence the parameter $z$ from each of the sheets  $\Pi_{l^{(k)}}$
may be carried across the interval $(\lambda_k,\lambda_{k+1})$ on 
the neighboring unphysical sheet $\Pi_{l^{(k+1)}}$
with $l_j^{(k+1)}=1- l_j^{(k)}$, $j\leq k,$
$l_{k+1}^{(k+1)}=1$  and  $l_j^{(k+1)}=0,$  $j>k+1$.
This means actually that  $z$ may be carried from 
$\Pi_0$ on all the two--body unphysical sheets 
$\Pi_{l^{(k+1)}}$ with
$l_j^{(k+1)}=0,$   $j >k+1$. For all this the parameter $z$ remains 
in the holomorphness 
domains $\Pi_{l^{(k+1)}}^{\rm (hol)}$ and does not visit the sheets 
$\Pi_{l^{(s)}}$ with $s>k+1$. By the principle of mathematical induction 
we conclude that the parameter $z$ may be carried really on all the 
two--body unphysical sheets. 

Proof is completed. 

Using results of Sec.~4 of the paper~\cite{M1} and Lemma~\ref{LPossibility},
one can prove the following important statement. 
\begin{theorem}\label{ThIter}\hspace*{-0.5em}{\sc .}
The iterations $\cQ\un(z)=\bigl((-\bt\bRo\Y)^n \bt \bigr)(z)$,\,\,
$n\geq 1$,   of absolute terms of the Faddeev equations (\ref{MFE})
admit in a sense of distributions over $\Osix$,
the analytical continuation on the domain $\Pilh$ of each unphysical 
sheet  $\Pi_l\subset\Re$. This continuation is described by 
the equalities
$
%
  \reduction{\cQ\un(z)}{\Pi_l}=\bigl((-\bt^l\bRo^l\Y)^n \bt^l \bigr)(z).
$
\end{theorem}
\begin{note}\label{NIterF}\hspace*{-0.5em}{\sc .}
{\rm The products $L_1\rJ_1\Psis\Y\cQ^{(m)}$, \,\,\,\, 
$\cQ^{(m)}\Y\Psi\rJt_1 L_1$, \,\,\,\,  $\tilde{L}_0\bJo\cQ^{(m)}$, \,\,\,\,
$\cQ^{(m)}\bJot\tilde{L}_0$,\,\,
\newline
$L_1\rJ_1\Psis\Y \cQ^{(m)}\Y\Psi\rJt_1 L_1$, \,\,\,\,
$\tilde{L}_0\bJo\cQ^{(m)}\bJot\tilde{L}_0,$ \,\,\,\, 
$L_1\rJ_1\Psis\Y\cQ^{(m)}\bJot\tilde{L}_0$\,\, and 
$\tilde{L}_0\bJo\cQ^{(m)}\Y\Psi\rJt_1 L_1$,\,\, $0\leq m < n$, 
arising at substitution of the relations (\ref{tltot})
and (\ref{R0ltot})  into  
$ \reduction{\cQ\un(z)}{\Pi_l}$, have to be understood  
in a sense of definitions from Sec.~4 of the paper~\cite{M1}.
}
\end{note}
\begin{note}\label{NDomainF}\hspace*{-0.5em}{\sc .}
{\rm Theorem~\ref{ThIter} means that one can 
pose the continued Faddeev equations  (\ref{MFEl}) only in the domains 
$\Pilh\subset\Pi_l$.
}
\end{note}
\section{\hspace*{-1em}.
                   REPRESENTATIONS FOR THE FADDEEV COMPONENTS \newline
                   OF THREE--BODY $T$--MATRIX }
\label{STRepres}

In the present section, using the Faddeev equations~(\ref{MFEl}) continued,
we shall obtain the representations for the matrix $M^l(z)$
in the domains $\Pilh$ of unphysical sheets $\Pi_l\subset\Re$.
The representations will be given in terms 
of the matrix $M(z)$ components taken on the physical sheet, 
or more precisely,  
in terms of the half--on--shell matrix $M(z)$ as well as 
the operators inverse to the truncated scattering 
matrices $S_l(z)$ and $\St_l(z)$.
As a matter of fact, the construction of the representations 
for  $M^l(z)$ consists in explicit 
 ``solving'' the continued Faddeev equations~(\ref{MFEl}) in the same way 
as in~\cite{MotTMF}, \cite{MotYaF} where the type~(\cite{M1}.3.2)
representations had been found for analytical continuation 
of the $T$--matrix in the multichannel scattering problem with binary channels. 
We consider derivation of the representations for 
$M^l(z)$ as a constructive proof of the existence  
(in a sense of distributions over $\Times_{\an=1}^3 \Osix$) 
of the analytical continuation of the matrix  $M(z)$ 
on unphysical sheets $\Pi_l$ of the surface $\Re$.

So, let us consider the Faddeev equations~(\ref{MFEl}) on the sheet 
$\Pi_l$ with  $l_0=0$ or $l_0=\pm 1$ and  
$l\bjd=0$ or $l\bjd=1,$ \,\, $\bnum$.
Using the expressions~(\ref{tltot}) for $\bt^l(z)$
and~(\ref{R0ltot}) for $\bRo^l(z)$,
we transfer all the summands including $M^l(z)$  
but not $\bJo$ and $\rJ_1$, to the left--hand part of Eqs.~(\ref{MFEl}).   
Making then a simple transformation based on the identity 
$
\bs_l^{-1}(z)=\hbIo -\bs_l^{-1}(z)\bJo(z)\bt(z)\bJot(z)A_0(z)L_0
$
we rewrite~(\ref{MFEl}) in the form 
\be
\label{MFEltr}
   (\bI + \bt \bRo\Y)M^l=
 \bt\left[ \bI-\Aol\bJot\bs_l^{-1}\bJo\bt - \Aol\bJo\Xol \right]-
\Phi\rJt_1 \Ail (\rJ_1\Phi^{*}+\Xil) 
\ee
where 
$
            \Aol(z)=L_0 A_0(z),
$\,\,\,
$
 \Ail(z)=L_1 A_1(z).
$
Besides we denote 
\be
\label{DefX0}
\begin{array}{l}
 \Xol=|L_0| \bs_l^{-1}\bJo(\bI-\bt\bRo)\Y M^l,\\
   \Xil=-L_1 \left[ \rJ_1\Phis\bRo+\Aol\rJ_1\Phis\bJot\bJo\right]\Y M^l.
\end{array}
\ee
It follows from Eq.~(\cite{M1}.3.5) that 
\be
\label{JFR0}
 \rJ_1\Phis\bRo=-\rJ_1\Psis.
\ee
Together with~(\ref{JFR0}) the equalities 
\be
\label{JFJtot}
\left( \rJ_1\Phis\bJot \right) (z)=0, \quad
\left( \bJo\Phi\rJt_1   \right) (z)=0,
\ee
take place being true in accordance with~(\ref{JFJ}) for all 
$z\in\bC\setminus(-\infty,\lambda_{\rm max}]$.

Using Eq.~(\ref{JFR0}) and first of Eqs.~(\ref{JFJtot}) 
one can rewrite $\Xil$ in the form 
\be
\label{DefX1}
\Xil=L_1 \rJ_1\Psis\Y M^l,
\ee
too. Note that the condition  $z\not\in(-\infty,\lambda_{\rm max})$
necessary for Eq.~(\ref{JFJtot}) to be valid,
does not touch the two--body unphysical sheets $\Pi_l$, $l_0=0,$
since in this case $\Aol(z)=0$
and consequently, the terms including the products $\bJot\bJo$,
are plainly absent in~(\ref{MFEltr}). 
Meanwhile the points 
$z\in(-\infty,\lambda_{\rm max}]$ were excluded from the three--body 
sheets $\Pi_l,$\,\, $l_0=\pm 1$, by definition.

Notice further that the operator $\bI+\bt\bRo\Y$ admits the 
explicit inversion in terms of  $M(z)$,
\be
\label{inv}
(\bI+\bt\bRo\Y)^{-1}=\bI-M\Y\bRo,
\ee
for all $z\in\Pi_0$ which do not belong to the discrete spectrum  
$\sigma_d(H)$ of the Hamiltonian $H$, and
\be
\label{invt}
(\bI-M\Y\bRo)\bt=M.
\ee
The equality~(\ref{inv}) is a simple consequence of the 
Faddeev equations~(\ref{MFE}) and the identity $\bRo\Y=\Y\bRo$.
The relation~(\ref{invt}) represents an alternative variant 
of these equations. Now, we can rewrite Eqs.~(\ref{MFEltr})
in the equivalent form 
\be
\label{Ml3in}
\begin{array}{c}
{M^l=M\left(\bI-\Aol\bJot\bs_l^{-1}\bJo\bt-\Aol\bJot\Xol\right)-}
                                                        \\
 - (\bI-M\Y\bRo)\Phi\rJt_1\Ail(\rJ_1\Phis+\Xil).
\end{array}
\ee
Eq.~(\ref{Ml3in}) means that the matrix $M^l(z)$
is expressed in terms of the quantities $\Xol(z)$ and $\Xil(z)$.
Main goal of the section consists really in 
presenting these quantities in terms of the matrix 
$M(z)$ considered on the physical sheet. 

To obtain for $\Xol$ and  $\Xil$ a closed system of equations 
we use the definitions~(\ref{DefX0}) and~(\ref{DefX1}) and 
act on the both parts of Eq.~(\ref{Ml3in})
by the operators $\bs_l^{-1}\bJo(\bI-\bt\bRo)\Y$
and $\rJ_1\Psis$. At this moment we use also the identities 
\be
\label{Help1}
[ \bI -\bt\bRo]\Y M=M_0 -\bt, \quad
[ \bI -\bt\bRo]\Y [\bI-M\Y\bRo]=[\bI-M_0\bRo]\Y
\ee
where $M_0=\Omt\Om M= (\bI+\Y) M.$
The relations~(\ref{Help1}) are another easily checked consequence 
of the Faddeev equations~(\ref{MFE}).
Along with Eq.~(\ref{Help1}) we apply second of the equalities
~(\ref{JFJtot}). As a result we come to the following system 
of equations for  $\Xol$ and $\Xil$:
\begin{eqnarray}
\Xol &=&|L_0| \bs_l^{-1}\bJo\left[ (M_0-\bt)(\bI-
\Aol\bJot\bs_l^{-1}\bJo\bt -
\Aol\bJot\Xol\right] -                              \nonumber \\
& & -|L_0| \bs_l^{-1}\bJo M_0\Y\Psi\rJt_1\Ail(\rJ_1\Phis+\Xil),
\phantom{MMMM}
\label{sys1}         \\
\Xil & =& L_1 \rJ_1\Psis\Y M(\bI-\Aol\bJot\bs_l^{-1}\bJot\bt-\Aol\bJot\Xol) -
\nonumber                                           \\
&& - L_1 \rJ_1\Psis\Y[\Phi+M\Y\Psi]\rJt_1\Ail(\rJ_1\Phis+\Xil).
\label{sys2}
\end{eqnarray}
It is convenient to write this system in the matrix form 
$\Btl\Xl=\Dtl$,\,\, $\Xl=(\Xol,\Xil)^\dagger$
with $\Btl=\{ \Btl_{ij}  \},\,\, i,j=0,1,$
the matrix consisting of operators standing at unknown  $\Xol$
and $\Xil$. By   
$\Dtl,$ \,\, $\Dtl=(\Dtl_0, \Dtl_1)^\dagger), $
we understand a column constructed of the absolute terms 
of Eqs.~(\ref{sys1}) and~(\ref{sys2}). Since 
$\bs_l=\hbIo+\Aol\bJo \bt\bJot$ we find 
$
\Btl_{00}  =  \bs_l^{-1} (\hbIo +\Aol\bJo M_0\bJot),
$
At the same time 
$
\Btl_{01}  =  |L_0| \bs_l^{-1}\bJo M_0\Y\Psi\rJt_1\Ail, \quad
\Btl_{10}  =  L_1 \rJ_1\Psis\Y M\bJot \Aol
$
and 
$
\Btl_{11}=\hIi+L_1 \rJ_1\Psis U\Psi\rJt_1\Ail
$
because 
$
\Y(\Phi+M\Y\Psi)=\Y\bv\Psi+\Y M\Y\Psi=(\Y \bv +\Y M\Y)\Psi =U\Psi
$
(see~\cite{M1}, Sec.~4).

The absolute terms look as 
\begin{eqnarray*}
\Dtl_0 & = & |L_0|\bs_l^{-1}[\bJo(M_0-\bt)(\bI-\Aol\bJot\bs_l^{-1}\bJo\bt)
-|L_0|\bJo M_0\Y\Psi\rJt_1\Ail\rJ_1\Phis],  \\
\Dtl_1 & = & L_1 \rJ_1\Psis\Y M(\bI-\Aol\bJot\bs_l^{-1}\bJo\bt)
- L_1 \rJ_1\Psis U\Psi\rJ_1\Ail\rJ_1\Phis.
\end{eqnarray*}

The operator $\bs_l(z)$, \,\, $l_0=\pm 1,$
has inverse one for all $z\in\cP_b$.
If $z\not\in Z_{\rm res}$ then $\bs_l^{-1}(z)$
is a bounded operator in $\hat{\cG}_0$. 
That is why, acting on the both parts of 
the first equation 
$ \Btl_{00}\Xol+\Btl_{01}\Xil=\Dtl_0$
of the system $\Btl\Xl=\Dtl$ by the operator $\bs_l$,
and not changing its second equation, we come to the equivalent system  
\be
\label{BXD}
\Bl\Xl=\Dl
\ee
with the (operator) matrix 
\be
\label{Bl}
\Bl=\left(\begin{array}{cc}
\hbIo+|L_0|\bJo M_0\bJot\Aol  & |L_0|\bJo M_0\Y \Psi\rJt_1\Ail   \\
L_1 \rJ_1\Psis\Y M\bJot\Aol & \hIi +L_1 \rJ_1\Psis U\Psi\rJt_1\Ail
\end{array}\right),
\ee
$\Bl(z):$
$\hat{\cG}_0\oplus\hat{\cH}_1\rightarrow \hat{\cG}_0\oplus\hat{\cH}_1 $,
and the absolute term $\Dl$ having the components 
$\Dl_0=\bs_l \Dtl_0$ and $\Dl_1=\Dtl_1$.
\begin{lemma}\label{LBlInv}\hspace*{-0.5em}{\sc .}
The operator $\left(\Bl(z)\right)^{-1}$
exists for all $z\in\Pi_l^{\rm (hol)}$ such that there exists the operator 
$S_l^{-1}(z)$ inverse to the truncated three--body scattering matrix 
$S_l(z)$ given by first of the equalities~(\ref{Slcut})
with $  L=\diag\{L_0,L_1\},$\,\,\, $\tL=\diag\{|L_0|,L_1\},$
and such that there exist the operators $[S_l(z)]^{-1}_{00}$
and $[S_l(z)]^{-1}_{11}$ inverse to 
$[S_l(z)]_{00}=\hat{I}_0+\rJo T\rJot A_0 L_0$
and $[S_l(z)]_{11}=\hat{I}_1+L_1 \rJ_1\Psis U\Psi\rJt_1 A_1 L_1,$
respectively. The components 
$\left[ \left(\Bl(z) \right)^{-1}\right]_{ij},$\,\, $ i,j=0,1,$
of the operator $\left(\Bl(z)\right)^{-1}$
admit the representation 
\begin{eqnarray}
\label{Y00}
\left[ \left(\Bl(z) \right)^{-1}\right]_{00} & = &
\hbIo-\Omt[S_l^{-1}]_{00}
\left\{|L_0|\rJo T_0 - [S_l]_{01}[S_l]_{11}^{-1}L_1\rJ_1\Psis\Y M\right\}
\bJot\Aol, \\
\label{Y01}
\left[ \left(\Bl(z) \right)^{-1}\right]_{01} & = & \Omt[S_l^{-1}]_{01},\\
\label{Y10}
\left[ \left(\Bl(z) \right)^{-1}\right]_{10} & = & -[S_l^{-1}]_{11}
L_1 \rJ_1\Psis\Y M\bJot\Aol
\left\{\hbIo -\Omt[S_l]^{-1}_{00}|L_0|\rJo T_0\bJot\Aol \right\},  \\
\label{Y11}
\left[ \left(\Bl(z) \right)^{-1}\right]_{11} & = & [S_l^{-1}]_{00}
\end{eqnarray}
with $T_0\equiv \Om M$.
\end{lemma}

Note that since $|L_0|$ and $\Aol$ are numbers turning into zero at $l_0=0$
simultaneously, the factors $|L_0|$ in (\ref{Y00}) and (\ref{Y10})
may be omitted.

{\sc Proof.} Let us find at the beginning, the components 
$\left[ \left(\Bl(z) \right)^{-1}\right]_{00}$
and $\left[ \left(\Bl(z) \right)^{-1}\right]_{10}$,
which will be denoted temporarily (for the sake of contracting the writing) 
by $Y_{00}$ and $Y_{10}$.
Using Eq.~(\ref{Bl}) we write the equation system for these components,
\begin{eqnarray}
\label{EQY00}
[\Bl]_{00}\,\, Y_{00}+[\Bl]_{01}\,\, Y_{10}&=&\hbIo\\
\label{EQY10}
[\Bl]_{10}\,\, Y_{00}+[\Bl]_{11}\,\, Y_{10}&=& 0
\end{eqnarray}
Eliminating the unknown $Y_{10}$  from the first equation~(\ref{EQY00}) 
with a help of~(\ref{EQY10}) we come to the following 
equation including the element $Y_{00}$ only, 
\be
\label{EY00}
\left\{ \hbIo+\Omt\left[|L_0|\rJo T_0\bJot\Aol -
[S_l]_{01}[S_l]^{-1}_{11} L_1 \rJ_1\Psis\Y M\bJot\Aol\right]   \right\}
Y_{00} = \hbIo.
\ee
At intermediate transforms we used the equality $M_0=\Omt T_0$.

The operator matrix in the left--hand part of Eq.~(\ref{EY00}) complementary to 
$\hbIo$, has three the same rows. Thus one can apply 
to Eq.~(\ref{EY00}) the inversion formula 
\be
\label{C123}
\left[ \hbIo+\Omt(C_1,\, C_2,\, C_3)\right]^{-1}= \hbIo -
\Omt\left[ \hIo+C_1+C_2+C_3\right]^{-1}(C_1,\, C_2,\, C_3),
\ee
which is true for a wide class of the operators $(C_1,\, C_2$ and $C_3)$.
A single essential requirement to 
$C_1,\, C_2$ and $C_3$ evidently, is the existence of  
$(\hat{I}_0+C_1+C_2+C_3)^{-1}$.

In the case concerned 
$$
C\bd(z)\equiv \left\{         |L_0|  \rJo T_{0\bn}\rJot
 -[S_l]_{01}[S_l]_{11}^{-1} L_1 \rJ_1\Psis\Y[M]\bd\rJot  \right\}\Aol
$$
where $[M]\bd$ is the 
$\beta$-th column of the matrix  $M$,
$[M]\bd=\Bigl(M_{1\bn},M_{2\bn},M_{3\bn}\Bigr)^\dagger$.
Thus
$$
\hIo+C_1+C_2+C_3=\hIo+\rJo T\rJot\Aol -[S_l]_{01}[S_l]_{11}^{-1}
\rJ_1\Psis U_0^{\dagger}\rJot\Aol \equiv
[S_l]_{00}-[S_l]_{01}[S_l]_{11}^{-1}[S_l]_{10}.
$$

Note that elements $\left[S_l^{-1}\right]_{ij}$,\,\, $i,j=0,1,$
of $S_l^{-1}$ may be present by the components  $\left[S_l\right]_{ij}$  as 
\begin{eqnarray}
\label{Si00}
\Slis_{00} &=&\left( \Sls_{00}-
\Sls_{01}\Sls_{11}^{-1}\Sls_{10}\right)^{-1}     \\
\label{Si11}
\Slis_{11} & = & \left( \Sls_{11}-
\Sls_{10}\Sls_{00}^{-1}\Sls_{01}\right)^{-1}     \\
\label{Si10}
\Slis_{10} & = & - \Sls_{11}^{-1}\Sls_{10}\Slis_{00}     \\
\label{Si01}
\Slis_{01} & = & - \Sls_{00}^{-1}\Sls_{01}\Slis_{11}
\end{eqnarray}
It follows from~(\ref{Si00}) that 
$\hIo+C_1+C_2+C_3=\left(\Slis_{00}\right)^{-1}$. Therefore
in the conditions of Lemma, the operator  
$(\hat{I}_0+C_1+C_2+C_3)^{-1}$ invertible. Now,  
a use of Eq.~(\ref{C123}) in~(\ref{EY00})
leads us immediately to the representation~(\ref{Y00}) for
$\left[ \left( \Bl\right)^{-1} \right]_{00}$.

When calculating $Y_{10}=\left[ \left( \Bl\right)^{-1} \right]_{10}$ 
we eliminate from the second equation~(\ref{EQY10}) vice versa, 
the quantity $Y_{00}$ using Eq.~(\ref{EQY00}).
For all this, we need to calculate the operator inverse to 
$\hbIo+\bJo M_0\bJot\Aol$.
Here, we apply again the relation~(\ref{C123}) and obtain that
\begin{eqnarray}
\left( \hbIo+|L_0|\bJo M_0\bJot\Aol \right)^{-1} & = &
\left( \hbIo+\Omt|L_0|\rJo T_0\bJot\Aol \right)^{-1}=
\nonumber\\
 & = & \hbIo-\Omt\Sls_{00}^{-1}|L_0|\rJo T_0\bJot\Aol.
\label{JM0J}
\end{eqnarray}

With a help of~(\ref{Slcut}) we can write the resulting equation for
$Y_{10}$ as
\begin{eqnarray}
\nonumber
\lefteqn{ \left\{ \Sls_{11}-
\Sls_{10}\Sls_{00}^{-1}\Sls_{01}\right\}Y_{10}=   }\\
\label{EY10}
& & = -\rJ_1\Psis\Y M\bJot\Aol
\left[ \hbIo+\bJo M_0\bJot\Aol \right]^{-1}.
\end{eqnarray}
According to~(\ref{Si11}) the expression in 
braces in the left--hand part of Eq.~(\ref{EY10}) coincides with 
$\Slis_{11}^{-1}$. Then from~(\ref{EY10}) we get 
immediately~(\ref{Y01}).

System of the equations 
\be
\label{EQY01}
[B^{(l)}]_{00}\,\, Y_{01} + [B^{(l)}]_{01}\,\, Y_{11} = 0
\ee
\be
\label{EQY11}
[B^{(l)}]_{10}\,\, Y_{01} + [B^{(l)}]_{11}\,\, Y_{11} = \hIi
\ee
for the components  $Y_{01}=[(B^{(l)})^{-1}]_{01}$
and $Y_{11}=[(B^{(l)})^{-1}]_{11}$
is solved analogously. Search for $Y_{11}$
is at all a simple problem because the use of 
the inversion formula~(\ref{JM0J}) 
for Eq.~(\ref{EQY01})  immediately gives
$
Y_{01}=\Omt\,\Sls_{00}^{-1}\, \Sls_{01}\, Y_{11}.
$
Substituting this $Y_{01}$ in~(\ref{EQY11}) we find 
$$
\left\{ \Sls_{11}-\Sls_{10}\Sls_{00}^{-1}\Sls_{01}\right\}\,
Y_{11}=\hIi.
$$
 Here, one can see in the left--hand part as in (\ref{EY10}),
the operator $\Slis_{11}^{-1}$. Inverting it, we come 
to Eq.~(\ref{Y11}).

When calculating the unknown $Y_{01}$, we begin with 
expressing by it the unknown $Y_{11}$. Using Eq.~(\ref{EQY11}) we get 
\be
\label{XY11}
Y_{11}=\Sls_{11}^{-1}\left(\hIi-L_1\rJ_1\Psi\Y M\bJo\Aol Y_{01}\right).
\ee
Substituting~(\ref{XY11}) into Eq.~(\ref{EQY01}) we obtain an equation 
with operator standing at $Y_{01}$, which may be inverted 
with a help of Eq.~~(\ref{C123}). Then we use also the chain of equalities 
$$
|L_0|\bJo M_0\Y\Psi\rJt_1\Ail=|L_0|\bJo\Omt\Om M\Y\Psi\rJt_1\Ail=
$$
$$
=\Omt|L_0|\rJo\Om M\Y\Psi\rJt_1\Ail=\Omt\Sls_{01},
$$
simplifying the absolute term as well as the summand 
in the left-hand part, engendered there 
due to~(\ref{XY11}) by the element $\left[\Bl\right]_{01}$.
Completing the transforms we find 
$$
Y_{01}=-\Omt\left\{
\Sls_{00}-\Sls_{01}\Sls_{11}^{-1}\Sls_{10}\right\}^{-1}
\Sls_{01}\Sls_{11}^{-1}.
$$
In view of~(\ref{Si01}), the expression standing 
after $\Omega^{\dagger}$ in the right--hand part 
of the last equation, coincides 
exactly with that for $\Slis_{01}$. Therefore finally, 
we obtain Eq.~(\ref{Y01}). Thus, all the components 
of the inverse operator $\left(\Bl\right)^{-1}$
have already been calculated. 

It follows from the representations~(\ref{Y00}) -- (\ref{Y11}) 
that  $\left(\Bl(z)\right)^{-1}$ exists 
for such $z\in\Pi_l^{\rm (hol)}$ 
that there exist the operators inverse to $S_l(z)$,\,
$\left[S_l(z)\right]_{00}$ and $\left[S_l(z)\right]_{11}$.

The lemma has been proved. 

Let us back to Eq.~(\ref{BXD}) and inverse in it, using the relations 
~(\ref{Y00}) -- (\ref{Y11}), the operator $\Bl(z)$. Thereby we find 
the unknowns $\Xol$ and $\Xil$ which express $M^l(z)$ 
[see Eq.~(\ref{Ml3in})]. 

When carrying a concrete calculation of 
$\Xol=\Blis_{00}\, \Dl_0+\Blis_{01}\Dl_1$
we use the relation 
$
    |L_0|\Blis_{00}\, \bJo\,  M_0 =\Omt|L_0|\Slis_{00}\, \rJo\, T_0
$
that can be checked with a help of~(\ref{Slcut}) and~(\ref{T3body}).
Along with the identity 
\be
\label{sjtl3}
\bJo\bt\left(\hbIo-\Aol\bJot\bs_l^{-1}\bJo\bt\right)=\bs_l^{-1}\bJo\bt, 
\ee
this relation simplifies essentially the transform of the product 
$ \Blis_{00}\Dl_0 $.
Besides when calculating $ \Xol$, we use the equalities~(\ref{JFJtot}).
As a result we find 
\be
\label{X0}
\Xol=\Omt\left\{ |L_0|\Slis_{00}\rJo T_0 +\Slis_{01}
L_1 \left(\rJ_1\Psis\Y M+\rJ_1\Phis \right) \right\}
-|L_0|\bs_l^{-1}\bJo\bt.
\ee

Now, to find 
$
%
 \Xil=\Blis_{10}\Dl_0+\Blis_{11}\Dl_1 ,
$
we observe additionally that the equality 
$
\left\{\hbIo-\Omt\Slis^{-1}_{00}\rJo T_0\bJot\Aol \right\}
\bJo M_0 =\Omt\Slis_{00}^{-1}\rJo T_0
$
simplifying the product $\Blis_{10}\, \Dl_0$, is valid. 
The final expression for $\Xil$ read as 
\be
\label{X1}
\Xil=L_1 \left\{ \Slis_{10}\, |L_0|\rJo\, T_0+
\Slis_{11}\, L_1\rJ_1\Psis\Y M-
\left( \hIi-\Slis_{11}\right) L_1\rJ_1\Psis\right\}.
\ee

To obtain now a representation for $M^l(z)$, one needs 
at the moment only to substitute 
the found expressions~(\ref{X0}) for $\Xol$
and~(\ref{X1}) for $\Xil$ in Eq.~(\ref{Ml3in}).
Carrying out series of simple but rather  
cumbersome transformations of Eq.~(\ref{Ml3in}) 
we come as a result to the statement analogous 
to Theorem~1 of Ref.~\cite{M1}  concerning analytical 
continuation of the two--body $T$--matrix.  The statement is following. 
\begin{theorem}\label{ThMlRepr}\hspace*{-0.5em}{\sc .}
The matrix $M(z)$ admits in a sense of distributions over 
$\Osix$, the analytical continuation in $z$ on the domains $\Pilh$
of unphysical sheets $\Pi_l$ of the surface $\Re$.
The continuation is described by 
\be
\label{Ml3fin}
M^l=M-\left(M\Omt\rJot,\,\,\,\,   \Phi\rJt_1+M\Y\Psi\rJt_1\right)
LA\,\, S_l^{-1}  \tL  \left(
\begin{array}{c}
\rJo\Om M \\ \rJ_1\Psis\Y M+ \rJ_1\Phis
\end{array}         \right)
\ee
where $S_l(z)$ is a truncated scattering matrix (\ref{Slcut}),\,\,
$L=\diag\{l_0,l_{1,1},...,$
$l_{1,n_1},$ $l_{2,1},...,$
$l_{2,n_2},$ $l_{3,1},...,$ $l_{3,n_3}\}$
and $\tL=\diag\{|l_0|,$ $l_{1,1},...,$
$l_{1,n_1},$ $l_{2,1},...,$
$l_{2,n_2},$ $l_{3,1},...,$ $l_{3,n_3}\}$.
Kernels of all the operators in the right--hand part of Eq.~(\ref{Ml3fin})
are taken on the physical sheet. 
\end{theorem}

Note that  $LA\,\, S_l^{-1}(z)  \tL = \tL [S^\dagger_l(z)]^{-1}\,\, AL$.
This means that the relations~(\ref{Ml3fin}) may be rewritten also 
in terms of the scattering matrices 
$S^\dagger_l(z)$.

\section{\hspace*{-1em}. ANALYTICAL CONTINUATION OF THE SCATTERING MATRICES}
\label{SSmxl}
Let $l=\left\{l_0,\, l_{1,1},...,l_{1,n_1},\right.$
 $l_{2,1},...,l_{2,n_2},$
 $\left. l_{3,1},...,l_{3,n_3}\right\}$
with certain $l_0$,\,\, $l_0=0$ or $l_0=\pm 1$
and $l\ajd$,\,\, $l\ajd=0$  or $l\ajd=+1$,\,\, $\anum$.
The truncated scattering matrices  $S_l(z):$
$\hat{\cH}_0\oplus\hat{\cH}_1\rightarrow\hat{\cH}_0\oplus\hat{\cH}_1$
and $\St_l(z):$
$\hat{\cH}_0\oplus\hat{\cH}_1\rightarrow\hat{\cH}_0\oplus\hat{\cH}_1$,
given by formulae (\ref{Slcut}),
are operator--valued functions of variable $z$ being holomorphic 
in the domain $\Pilh$ of the physical sheet $\Pi_0$.
At $l_0=1$ and $l\ajd=1$,\, $\anum$,
these matrices coincide with the respective 
total three--body scattering matrices:
$S_l(z)=S(z),$ \, $\St_l(z)=\St(z).$

We describe now the analytical continuation of $S_{l'}(z)$ and $\St_{l'}(z)$
with a certain multi--index $l'$
on unphysical sheets $\Pi_{l}\in\Re$.
We shall base here on the representations (\ref{Ml3fin})
for $\reduction{M(z)}{\Pi_{l}}$. As mentioned above, our goal is 
to find the explicit representations for 
$\reduction{S_l(z)}{\Pi_{l'}}$
and  $\reduction{\St_l(z)}{\Pi_{l'}}$ again in terms of the 
physical sheet.

First of all, we remark that the function $A_0(z)$ is 
univalent. It looks as $A_0(z)=-\pi i z^2$ on all the sheets $\Pi_l$. 
At the same time after continuing from $\Pi_0$ on $\Pi_l$, 
the function $A\bjd(z)=-\pi i\sqrt{z-\lambda\bjd}$
keeps its form if only $l\bjd=0$.
If $l\bjd=1$ this function turns into $A'\bjd(z)=-A\bjd(z)$.
Analogous inversion takes (or does not take) place 
for arguments $\hP$, $\hP'$, $\hp\ad$ and $\hp'\bd$
of kernels of the operators 
$\rJo \Om M\Omt\rJot$,\,\, $\rJo \Om M\Y\Psi\rJt_1$,\,\,
$\rJ_1\Psis\Y M\Omt\rJot$ and $\rJ_1\Psis(\Y\bv+\Y M\Y)\Psi\rJt_1$, too.
Remember that on the physical sheet $\Pi_0$, 
the action of $\rJo(z)$\,\,  ($\rJot(z)$) transforms 
$P\in\Rs$ in $\sqrt{z}\hP$ \,\, ($P'\in\Rs$ in $\sqrt{z}\hP'$). 
At the same time, 
$p\ad\in\Rt$ \, ($p'\bd\in\Rt$)\, turns under $\rJ_{\an,i}(z)$\,\,
($\rJt_{\bn,j}(z)$)\,\, into 
$\sqrt{z-\lambda_{\an,i}}\, \hp\ad$ \,\,
($\sqrt{z-\lambda_{\bn,j}}\, \hp'\bd$).
That is why  we introduce the operators 
$\cE(l)=\diag\{\cE_0,\,\cE_1\}$ 
where $\cE_0$ is the identity operator in $\hat{\cH}_0$ if 
$l_0=0$, and $\cE_0$, the inversion, 
$(\cE_0 f)(\hP)=f(-\hP)$ if $l_0=\pm 1$. Analogously
$\cE_1(l)=\diag\{ \cE_{1,1},...,\cE_{1,n_1};$ $\cE_{2,1},...,\cE_{2,n_2};$
$\cE_{3,1},...,\cE_{3,n_3}\}$ where 
$\cE\bjd$ is the identity operator in $\hat{\cH}^{(\bn,j)}$ if
$l\bjd=0$, and $\cE\bjd$, the inversion 
$(\cE\bjd f)(\hp\bd)=f(-\hp\bd)$ if $l\bjd=1$.
By  $\re_1(l)$ we denote the diagonal matrix 
$\re_1(l)=\diag\{
\re_{1,1},...,\re_{1,n_1};$ $\re_{2,1},...,\re_{2,n_2};$
$\re_{3,1},...,\re_{3,n_3}\}$
with elements  $\re\bjd=1$ if $l\bjd=0$ and $\re\bjd=-1$ if $l\bjd=1$.
Let  $\re(l)=\diag\{ \re_0, \re_1 \}$
where $\re_0=+1$.
\begin{theorem}\label{ThS3lRepr}\hspace*{-0.5em}{\sc .}
If there exists a path on the surface $\Re$ such that 
at moving by it from the domain $\Pi^{\rm (hol)}_{l'}$
on $\Pi_0$ to the domain $\Pi^{\rm (hol)}_{l'}\bigcap\Pi^{\rm (hol)}_{l'l}$
on $\Pi_{l}$, the parameter $z$ stays on 
intermediate sheets $\Pi_{l''}$ always in the domains 
$\Pi^{\rm (hol)}_{l'}\bigcap\Pi^{\rm (hol)}_{l''l'}$,
then the truncated scattering matrices $S_{l'}(z)$ and $\St_{l'}(z)$
admit the analytical continuation in $z$
on the domain $\Pi^{\rm (hol)}_{l'}\bigcap\Pi^{\rm (hol)}_{l'l}$ 
of the sheet $\Pi_{l}$.
The continuation is described by 
\begin{eqnarray}
\reduction{S_{l'}(z)}{\Pi_{l}} & = & \cE(l)\left[ \hbI+
\tL' \hat{\cT} L'\,\, A \re(l) -
\tL' \hat{\cT} L\,\, A\, S^{-1}_{l}\,\, \tL \hat{\cT} L'\,\, A\re(l)
\right] \cE(l),
\label{Slfin}                               \\
\reduction{\St_{l'}(z)}{\Pi_{l}} & = & \cE(l)\left[ \hbI+
 \re(l) A \,\, L' \hat{\cT} \tL'  -
\re(l) A\,\, L \hat{\cT} \tL\,\, [\St_{l}]^{-1}\, A \,\, L \hat{\cT} \tL'
\right] \cE(l),
\label{Stlfin}
\end{eqnarray}
where $L'=\left\{l'_0,\, l'_{1,1},...,l'_{1,n_1},\right.$
 $l'_{2,1},...,l'_{2,n_2},$
 $\left. l'_{3,1},...,l'_{3,n_3}\right\}$ and 
$\tL'=\left\{|l'_0|,\, l'_{1,1},...,l'_{1,n_1},\right.$
 $l'_{2,1},...,l'_{2,n_2},$
 $\left. l'_{3,1},...,l'_{3,n_3}\right\}$.
\end{theorem}
{\sc Proof.} We give the proof for example of $S_{l'}(z)$.
Using the definition~(\ref{T3body}) of the operator $\cT(z)$
we rewrite $S_{l'}(z)$ in the form 
$$
S_{l'}(z)=\hbI+\tL'\left[        \left(\begin{array}{c}
\rJo \Om     \\  \rJ_1\Psis\Y   \end{array}\right)
 M \left(   \Omt\rJot\, ,\,\,\,\,    \Y\Psi\rJt_1  \right)
 + \left(\begin{array}{cc}
 0      &      0     \\
 0      &   \rJ_1\Psis\Y\bv\Psi\rJt_1
\end{array}\right)
\right]\, L' A.
$$
Note that when continuing on the sheet $\Pi_{l''}$, the operators 
$\rJo(z),$\, $\rJot(z),$\, $\rJ_1(z)$ and  $\rJt_1(z)$
turn into  $\cE_0(l'')\rJo(z),$\,
$\rJot(z)\cE_0(l''),$\, $\cE_1(l'')\rJ_1(z)$
and  $\rJt_1(z)\cE_1(l'')$, respectively. At the same time 
the matrix--function $A(z)$ turns into $A(z)\re(l'')$. 
Then using Theorem~\ref{ThMlRepr}, for the domains  
$\Pi^{\rm (hol)}_{l'}\bigcap\Pi^{\rm (hol)}_{l'l''}$
of intermediate sheets $\Pi_{l''}$ we have
\begin{eqnarray}
\nonumber
\lefteqn{\reduction{S_{l'}(z)}{\Pi_{l''}}=\hbI+
\cE(l'')\tL'\hat{\cT}L'\cE(l'') A\re(l'') -  }  \\
\label{S3l}
&& -\cE(l'')\tL' \left( \begin{array}{c}
 \rJo\Om    \\      \rJ_1\Psis\Y
\end{array} \right)
\left( M\Omt\rJot\, ,\,\,\, [\bv+M\Y]\Psi\rJt_1  \right)
L''A\, S_{l''}^{-1}\times \\
\nonumber && \times \tL''\left( \begin{array}{c}
 \rJo\Om M   \\      \rJ_1\Psis[\bv+\Y M]
\end{array} \right)
\left( \Omt\rJot\, ,\, \Y\Psi\rJt_1  \right)L' \cE(l'') A \re(l'')
\end{eqnarray}
where the summand following immediately by $\hbI$,
is engendered by the term $M(z)$ of the right--hand part of~(\ref{Ml3fin}).
The last summand of~(\ref{S3l}) is originated from the second 
summand of~(\ref{Ml3fin}).

In view of~(\ref{JFJtot}) we have 
$\rJ_1\Psis\bv\Omt\rJot=\rJ_1\Phis\rJot\Omt=0$.  Analogously, 
$\rJo\Om\bv\Psi\rJt_1$ equals to zero, too. Thus, taking into account 
(\ref{T3body}) we find
\be
\label{S3ll}
\reduction{S_{l'}(z)}{\Pi_{l''}}=\hbI+
\cE(l'')\,\, \tL'\hat{\cT}L'\,\, \cE(l'') A\re(l'') -
\cE(l'')\,\, \tL'\hat{\cT}L''\,\, A\, S_{l''}^{-1}\,\,
\tL''\hat{\cT}L'\,\, \cE(l'')\, A\re(l'').
\ee
By the supposition, the parameter $z$ moves along such a path 
that on the sheet $\Pi_{l''}$ it is situated in the domain 
$\Pi^{\rm (hol)}_{l'}\bigcap\Pi^{\rm (hol)}_{l'l''}. $
In this domain, the operators 
$(\tL'\hat{\cT}L')(z)$,\,\, $(\tL'\hat{\cT}L'')(z)$ and 
$(\tL''\hat{\cT}L')(z)$ are defined and depending on $z$ analytically.
Consequently, the same may be said also 
about the function $\reduction{S_{l'}(z)}{\Pi_{l''}}$. 
In equal degree, this statement is related to the sheet 
$\Pi_{l}$. Replacing the values of multi--index $l''$ in the 
representations  (\ref{S3l}) and (\ref{S3ll}) with $l$,
we come to the assertion of theorem for $\reduction{S_{l'}(z)}{\Pi_{l}}$. 
Truth of the representations 
(\ref{Stlfin}) for  $\reduction{\St_{l'}(z)}{\Pi_{l}}$
is established in the same  way. 

The proof is completed.
\begin{note}\label{NSlPil}\hspace*{-0.5em}{\sc .}
{\rm If $l_0=0$ then the representation (\ref{Slfin})
for the analytical continuation of $S_l(z)$
on the (its ``own'') sheet $\Pi_l$ acquires the simple form 
[cf.~(\cite{M1}.3.6)],
$$
\reduction{S_l(z)}{\Pi_{l}}=\cE(l)\left[ \hbI +\re(l) -
S_l^{-1}(z)\re(l)  \right]\cE(l)=\cE(l)\, S_l^{-1}(z)\, \cE(l).
$$
Just so 
$\reduction{\St_l(z)}{\Pi_{l}}=\cE(l)\, [\St_l(z)]^{-1}\, \cE(l).$
}
\end{note}

\section{\hspace*{-1em}. REPRESENTATIONS FOR ANALYTICAL CONTINUATION 
\newline
                          OF RESOLVENT }
\label{SResolvl}
The resolvent $R(z)$
of the Hamiltonian $H$ for three--body system concerned 
is expressed by $M(z)$ according to Eq.~(\ref{RMR}). 
As we established, kernels of all the operators included 
in the right--hand part of~(\ref{RMR}) admit in a sense
of distributions over  $\Osix$, the analytical continuation 
on the domains $\Pilh$ of unphysical sheets $\Pi_l\subset\Re$.
So, such continuation is admitted as well for the kernel 
$R(P,P',z)$ of $R(z)$.
Moreover, there exists an explicit representation for this continuation 
analogous to the representation (\cite{M1}.3.7) for two--body 
resolvent. 
\begin{theorem}\label{ThResolvl}\hspace*{-0.5em}{\sc .}
The analytical continuation, in a sense of distributions over $\Osix$,  
of the resolvent $R(z)$ on the domain $\Pilh$ of unphysical sheet 
$\Pi_l\subset\Re$ is described by  
\begin{eqnarray}
\nonumber
\lefteqn{\reduction{R(z)}{\Pi_l}=R+} \\
\label{R3l}
 & & +\bigl( [I-RV]\rJot\, ,\,\,\,\, \Om[\bI-\bRo M\Y]\Psi\rJt_1 \bigr)
LA S_l^{-1} \tL \left(
\begin{array}{c}  \rJo[I-VR]   \\     \rJ_1 \Psis[\bI-\Y M\bRo]\Omt
\end{array} \right).
\end{eqnarray}
Kernels of all the operators present in the right--hand part 
of Eq.~(\ref{R3l}) are taken on the physical sheet.
\end{theorem}
{\sc Proof.} For analytical continuation $R^l(z)$
of the kernel  $R(P,P',z)$ of $R(z)$ on the sheet $\Pi_l$ 
we have according to~(\ref{RMR}),  
\be
\label{RMRl}
  R^l(z)=R_0^l(z)-R_0^l(z) \Om M^l(z) \Omt R_0^l(z).
\ee
For $M^l(z)$ we have found already the representation~(\ref{Ml3fin}).
Since $R^l_0=R_0+L_0 A_0 \rJot\rJo$ we can rewrite Eq.~(\ref{RMRl})
in the form 
\begin{eqnarray}
\nonumber
R^l & = & R_0-R_0\Om M^l\Omt R_0
+A_0 L_0 \rJot\left( \hIo-\rJo\Om M^l\Omt\rJot L_0 A_0 \right)\rJo - \\
\label{Rlini}
 & & -A_0 L_0 \rJot\rJo\Om M^l\Omt R_0 -R_0\Om M^l\Omt\rJot\rJo L_0 A_0.
\end{eqnarray}
Consider separately the contributions of each summand of (\ref{Rlini}). 
Doing this we shall use the notations 
$$
\bB=\left( \Om M\Omt\rJot\, ,\,\,\,\, \Om M\Y\Psi\rJt_1+\Om\Phi\rJt_1 \right)
\mbox{ and }
\bBt=\left(
\begin{array}{c}   \rJo\Om M\Omt \\
                 \rJ_1\Psis\Y M\Omt +\rJ_1\Phis\Omt
\end{array}
\right).
$$

It follows from~(\ref{Ml3fin}) that
$
\Om M^l\Omt =\Om M \Omt -\bB LA\,\, S_l^{-1}\,\, \tL\bBt.
$
Hence two first summands of~(\ref{Rlini}) give together 
$$
R_0-R_0\Om M\Omt R_0 + R_0\bB L\, A  S_l^{-1} \,\tL\bBt R_0=
R + R_0\bB L\, A S_l^{-1}\, \tL\bBt R_0.
$$

Transforming the third term of~(\ref{Rlini}) we use again 
the representation~(\ref{Ml3fin}). We find 
\begin{eqnarray*}
\lefteqn{\rJo\Om M^l\Omt\rJot L_0 A_0  =
\hat{\cT}_{00} L_0 A_0- \left(\hat{\cT}_{00}\, ,\,\,\,\,\hat{\cT}_{01}\right)
L\, A S_l^{-1}\,\tL\left(    \begin{array}{c}
   \hat{\cT}_{00}   \\  \hat{\cT}_{10}    \end{array} \right)  L_0 A_0 =}\\
 & = & \omo \hat{\cT} L \, A \omos -
 \omo \hat{\cT} L\,\, A\, S_l^{-1}\, \tL\hat{\cT} L\, A\omos=
\omo\hat{\cT} L\, A\left(  \hbI-S_l^{-1} \tL\hat{\cT} L\,\, A
\right)\tL\omos
\end{eqnarray*}
where  $\omo$ stands for the projector acting from  
$\hat{\cH}_0\oplus\hat{\cH}_1$
to  $\hat{\cH}_0$ as 
$\omo\left( \begin{array}{c} f_0 \\ f_1 \end{array} \right)=f_0$,\,\,
$f_0\in \hat{\cH}_0$,\,\, $f_1\in\hat{\cH}_1$. By $\omos$ we understand 
as usually the operator adjont to $\omo$.
So far as $S_l=\hbI+ \tL\hat{\cT} L\, A$ we have 
$
    \hbI-S_l^{-1}\tL\hat{\cT} L\, A=
$
$
   S_l^{-1}\left( \hbI+ \tL\hat{\cT} L\, A-\tL\hat{\cT} L\, A\right)
$
$
   = S_l^{-1}.
$
Taking in account that  $L=L\cdot\tL$ we find 
$$
L_0 A_0 (\hIo-\rJo\Om M^l\Omt\rJot L_0 A_0) =
\omo A L(\hbI - \tL\hat{\cT} LA S_l^{-1}) \tL \omos=
\omo L A\, S_l^{-1}\, \tL\omos.
$$
This means that the third term of~(\ref{Rlini}) may be present as 
$\rJot\omo L\, A S_l^{-1}\,\tL\omos.$

When studying the fourth summand of (\ref{Rlini})
we begin with transforming the product 
$A_0 L_0\rJo \Om M^l \Omt$ to more convenient form. 
It follows from (\ref{Ml3fin}) that  
$$
A_0 L_0\rJo\Om M^l\Omt=A_0 L_0\rJo\Om M\Omt -
A_0 L_0\left(\hat{\cT}_{00}\, ,
\,\,\,\,    \hat{\cT}_{01}\right)
LA\, S_l^{-1}\,\tL\bBt.
$$
In view of $A_0 L_0\rJo\Om M\Omt=\omo A L\bBt$ and 
$
A_0 L_0\left(\hat{\cT}_{00},\hat{\cT}_{01}\right)L =
\omo AL\hat{\cT} L
$
we have 
$$
A_0 L_0\rJo \Om M^l\Omt=\omo \left( AL-
A\, L\hat{\cT}L\, A S_l^{-1} \tL\right) \bBt=\omo L\, A S_l^{-1}\,\tL\bBt.
$$
Analogously, in the fifth term of (\ref{Rlini}),
$
\Om M^l\Omt\rJot L_0 A_0  =
\bB\tL \bigl[S_l^{\dagger}\bigr]^{-1}A\, L \omos =
 \bB L\, A S_l^{-1}\, \tL\omos.
$
Thus two last summands of (\ref{Rlini}) give together 
$
-\rJot\omo L\, A S_l^{-1}\,\tL\bBt R_0 -
$
$
R_0\bB L\, A S_l^{-1}\,\tL\omos\rJo.
$
Substituting the expressions obtained into Eq.~(\ref{Rlini}) 
we find
$$
R^l=R+\left(\rJot\omo-R_0\bB\right)L\, A S_l^{-1}\tL
\left(\omos\rJo-\bBt R_0\right).
$$
Taking into account the definitions of $\bB$ and  $\bBt$ as well as 
the fact that  $R_0\Om M\Omt=RV$, \,\,
$\Omt M\Om R_0=VR$
(see~\cite{Faddeev63}, ~\cite{MF})
and  $R_0\Om\Phi\rJ_1=-\Om\Psi\rJ_1,$\,\,
$\rJ_1\Phis\Omt R_0=-\rJ_1\Psis\Omt$,
we come finally to Eq.~(\ref{R3l}) and this completes the proof. 
\section{\hspace*{-1em}. ON USE OF THE DIFFERENTIAL FADDEEV EQUATIONS\newline
	                 FOR COMPUTATION OF THREE--BODY RESONANCES }
\label{SNumerMethod}
As follows from the representations (\ref{Ml3fin}), (\ref{Slfin})
and  (\ref{R3l}), the matrices  $\reduction{M(z)}{\Pi_l}$,\,\,
$\reduction{S_{l'}(z)}{\Pi_l}$ and 
the Green function $\reduction{R(z)}{\Pi_l}$
may have poles at points belonging to the discrete spectrum $\sigma_d(H)$ 
of the Hamiltonian $H$. Nontrivial singularities 
of $\reduction{M(z)}{\Pi_l}$,\,\,
$\reduction{S_{l'}(z)}{\Pi_l}$ and  $\reduction{R(z)}{\Pi_l}$
correspond to those points $z\in\Pi_0\bigcap\Pilh$ where 
the inverse truncated scattering matrix 
$[S_l(z)]^{-1}$ (or $[\St_l(z)]^{-1}$ and it is the same) does not exist 
or where it represents an unbounded operator. 
The points $z$ where  $[S_l(z)]^{-1}$ does not exist, 
engender poles for  $\reduction{M(z)}{\Pi_l}$,\,\,
$\reduction{S_{l'}(z)}{\Pi_l}$ and  $\reduction{R(z)}{\Pi_l}$. 
Such points are called (three--body) resonances. 

The necessary and sufficient condition~\cite{ReedSimon} 
of irreversibility of the operator $S_l(z)$ for given $z$ consists in 
existence of non--trivial solution 
$\cA^{\rm (res)}\in\hat{\cH}_0\oplus\hat{\cH}_1$ to the equation 
\be
\label{ZeroSl}
    S_l(z)\cA^{\rm (res)}=0.
\ee
Investigation of this equation may be carried out 
on the base of the  results of Sec.~4 of the paper~\cite{M1} 
concerning properties of kernels of the operator 
$\hat{\cT}(z)$. In view of the space shortage we 
postpone this investigation for another paper. 

The equation (\ref{ZeroSl}) may be applied for a practical 
computations of resonances situated 
in the domains $\Pilh\subset\Pi_l$. The resonances have to be considered  
as those values of $z\in\Pi_0\bigcap\Pilh$ for which the operators  
$S_l(z)$ and $\St_l(z)$ have eigenvalue zero. 

Elements of the scattering matrices  $S_l(z)$ and $\St_l(z)$ 
are expressed in terms of the amplitudes (continued in the energy 
$z$ on the physical sheet) for different processes taking 
place in the three--body system under consideration. 
Respective formulae~\cite{MF} written  for the components of 
$\hat{\cT}$, read as 
$$
\begin{array}{lcl}
\hat{\cT}_{\an,j;\bn,k}(\hp\ad,\hp'\bd,z) &=&
C_0^{(3)}(z)\,\,{\cA}_{\an,j;\bn,k}(\hp\ad,\hp'\bd,z) ,\\
\hat{\cT}_{\an,j;0}(\hp\ad,\hP',z) &=&
C_0^{(3)}(z)\,\,\cA_{\an,j;0}(\hp\ad,\hP',z) \\
\hat{\cT}_{0;\bn,k}(\hP,\hp'\bd,z) &=&
C_0^{(6)}(z)\,\,\cA_{0;\bn,k}(\hP,\hp'\bd,z),\\
\hat{\cT}_{00}(\hP,\hP',z) &=&
C_0^{(6)}(z)\,\,\cA_{00}(\hP,\hP',z)
\end{array}
$$
with 
$
C_0^{(N)}(z)=-\Frac{ {\rm e}^{ i\pi(N-3)/4 }  }
{ 2^{(N-1)/2}\,\, \pi^{(N+1)/2}\,\, z^{(N-3)/4} }
$
where for the function $z^{(N-3)/4}$ one takes the main branch.  
The functions  $\cA_{\an,j;\bn,k}$ represent amplitudes of elastic 
($\an=\bn;$ $j=k$)
or inelastic ($\an=\bn;$ $j\neq k$)
scattering and rearrangement ($\an\neq\bn$)  for the 
($2\rightarrow 2,3)$ process, in the initial state of which 
the pair subsystem is in the $k$-th bound state 
and the complementary particle is asymptotically free.
The function $\cA_{0;\bn,k}$
represents for this process, a breakup amplitude of the system into three 
particles.  The amplitudes  $\cA_{\an,j;0}$ and $\cA_{00}$
correspond to processes respectively,  ($3\rightarrow 2)$
and ($3\rightarrow 3)$ in the state where initially, all three particles 
are asymptotically free. 
Remember that contributions to $\cA_{00}$ from  the single and double 
rescattering represent  singular distributions (см.~\cite{M1}).

Describing in Sec.~4 of the paper~\cite{M1} the analytical 
properties in variable $z$ and the smoothness properties in angular 
variables  $\hP$ or  $\hp\ad$ and $\hP'$ or $\hp'\bd$,  of the matrix 
$\hat{\cT}$ kernels we have described thereby as well 
the properties of the amplitudes  $\cA(z)$.

To search for the amplitudes $\cA(z)$ continued on the physical sheet, 
on can use e.g., the formulation~\cite{MF}, \cite{EChAYa} 
of three--body scattering problem based on the Faddeev differential equations 
for components of the scattering wave functions considered in 
the coordinate space. It is necessary only to come in this formulation, 
to complex values of energy  $z$.  The square roots 
$z^{1/2}$ and $(z-\lambda\ajd)^{1/2}$, \, $\anum$, presenting in the formulae 
of~\cite{MF}, \cite{EChAYa} determining asymptotical boundary conditions 
at the infinity, have to be considered as the main branches of
$\sqrt{z}$ and $\sqrt{z-\lambda\ajd}$.  Solving the Faddeev differential 
equations with such conditions one finds really the analytical 
continuation on the physical sheet for the wave functions and consequently, 
for the amplitudes  $\cA(z)$.
Knowing the amplitudes $\cA(z)$, one can construct a necessary 
truncated scattering matrix $S_l(z)$ and then find those values of 
$z$ for which there exits a nontrivial solution $\cA^{\rm (res)}$ 
to Eq.~(\ref{ZeroSl}). As mentioned above these values of  
$z$ represent the three--body resonances on respective sheet $\Pi_l$.
\medskip 

Concluding the paper we make the following remark. 

It is well known~\cite{MF} that a generalization of the 
Faddeev equations~\cite{Faddeev63} on the case 
of systems with arbitrary number of particle is represented 
by the Yakubovsky equations~\cite{Yakubovsky}.  The latter 
have the same structure as the Faddeev equations. 
Thus the scheme used in the present paper, may be applied as well 
to construction of the type (\ref{Ml3fin}), (\ref{Slfin}) and (\ref{R3l})  
explicit representations for analytical continuation 
of the $T$-- and scattering matrices and resolvent 
on unphysical part of the energy Riemann surface in the $N$--body 
problems with arbitrary $N$.
\section{\hspace*{-1em}. Acknowledgements}
The author is grateful to Professors~S.Albeverio, 
V.B.Belyaev, K.A.Ma\-ka\-rov and S.L.Ya\-kov\-lev for 
fruitful discussions. Also, the author is indebted to 
Prof.~P.Exner for useful remark.

\end{document}